\newcommand\has{\hat{S}}
\newcommand\deq{\delta Q_{rev}}
\newcommand\pa{\partial}

\newcommand\beq{\begin{equation}}
\newcommand\eeq{\end{equation}}
\newcommand\beqnl{\begin{eqnarray}}
\newcommand\beqna{\begin{eqnarray*}}
\newcommand\eeqna{\end{eqnarray*}}
\newcommand\eeqnl{\end{eqnarray}}

 \def\NN{\hbox{\sf I\kern-.13em\hbox{N}}}
 \def\HH{\hbox{\sf I\kern-.13em\hbox{H}}}
 \def\DD{\hbox{\sf I\kern-.13em\hbox{D}}}
 \def\RR{\hbox{\sf I\kern-.14em\hbox{R}}}
 \def\CC{\hbox{\sf I\kern-.44em\hbox{C}}}
 \def\ZZ{{\hbox{\sf Z\kern-.43emZ}}}
 \def\QQ{\hbox{\sf C\kern -.48emQ}}
 \def\Cc{\hbox{\sf C\kern -.47em {\raise .48ex \hbox{$\scriptscriptstyle |$}}
   \kern-.5em {\raise .48ex \hbox{$\scriptscriptstyle |$}} }}
 \def\Qq{\hbox{\sf Q\kern -.57em {\raise .48ex \hbox{$\scriptscriptstyle |$}}
   \kern-.55em {\raise .48ex \hbox{$\scriptscriptstyle |$}} }}

\documentstyle[prd,aps,amsfonts]{revtex}
\begin{document}
\draft

\vskip1pc

\title{Quasi-Homogeneous Thermodynamics and Black Holes}
\author{F. Belgiorno\footnote{E-mail address: belgiorno@mi.infn.it}}
\address{Dipartimento di Fisica, Universit\`a degli Studi di Milano,
%\affiliation{Dipartimento di Fisica, Universit\`a degli Studi di Milano, 
Via Celoria 16, 20133 Milano, Italy}
\date{\today}

\maketitle

\begin{abstract}

We propose a generalized thermodynamics in which  
quasi-homogeneity of the thermodynamic potentials plays a 
fundamental role. This thermodynamic formalism arises from 
a generalization of the approach presented in paper [1], 
and it is based on the requirement that quasi-homogeneity is a 
non-trivial symmetry for the Pfaffian form $\deq$. 
It is shown that quasi-homogeneous thermodynamics  
fits the thermodynamic features of at least 
some self-gravitating systems.  
We analyze how quasi-homogeneous thermodynamics 
is suggested by black hole thermodynamics. Then, 
some existing results involving self-gravitating systems 
are also shortly discussed in the light of 
this thermodynamic framework. The consequences of the lack of 
extensivity are also recalled. We show that 
generalized Gibbs-Duhem equations arise as a consequence of 
quasi-homogeneity of the thermodynamic potentials. 
An heuristic link between this generalized 
thermodynamic formalism and the thermodynamic limit is also 
discussed.

\end{abstract}

%\maketitle

\vskip 0.7truecm
%\maketitle

\newpage

\section{introduction}
\label{qotde}

In paper \cite{belhom} we have shown that, by choosing the extensive 
variables $(U,V,X^1,\ldots,X^n)$ as independent variables, and by 
postulating that the integrable Pfaffian form $\delta Q_{rev}$ representing 
the infinitesimal heat exchanged reversibly is homogeneous of 
degree one and that the homogeneity symmetry is non-trivial [for the 
definition of non-trivial symmetry see the next 
section], 
it is possible to find immediately an integrating factor for 
$\delta Q_{rev}$. In fact, 
if $Y=U\; \pa_U+V\; \pa_V +\sum_i\; X^i\; \pa_{X^i}$ is the Liouville 
vector field associated with the homogeneity symmetry, one 
has that an integrating factor is given by $f=\deq(Y)=i_Y \deq\not \equiv 0$ 
(see \cite{belhom} and references therein). 
The entropy $S(U,V,X^1,\ldots,X^n)$, which represents 
the fundamental equation in the entropy representation, is then   
straightforwardly recovered. 
The framework presented in paper \cite{belhom} 
can be generalized in the 
following sense. A generalized thermodynamics where $\delta Q_{rev}$ 
is a quasi-homogeneous integrable Pfaffian form is proposed. 
Quasi-homogeneity is realized to be a property which characterizes 
the behavior of thermodynamic functions like e.g. the entropy in the 
black hole case, in some statistical mechanical models 
involving non-relativistic matter interacting via the Newtonian 
potential and also in the case of self-gravitating radiation. 
When gravity plays an important role the standard extensivity 
property of thermodynamics does 
not hold; nevertheless, one 
can still find a form of thermodynamics in which 
each thermodynamic variable follows a power scaling law where 
the power can be different from one or zero, i.e., the variables are 
allowed to be neither extensive nor intensive. Black hole thermodynamics 
is the most evident and special thermodynamics belonging to this framework.
\\ 
\\
The plan of the paper is the following. In sect. \ref{qopf} 
we give some definitions and then in sect. \ref{general}, 
we propose a theoretical 
framework for quasi-homogeneous thermodynamics, by generalizing 
the standard thermodynamics case \cite{belhom}. In sects. 
\ref{bhtd},\ref{hnt} and \ref{geon} examples displaying a  
quasi-homogeneous thermodynamics are discussed. 
In sect. \ref{bhtd}, the role of quasi-homogeneity in 
black hole thermodynamics is shown; 
in sect. \ref{hnt}, 
we discuss a model of fermionic non-relativistic matter with 
Newtonian interaction which displays a quasi-homogeneous behavior. 
In sect. \ref{geon}, the general relativistic case of 
thermal geons and self-gravitating radiation are discussed. 
Sect. \ref{nonh} 
concerns the discussion of the physical consequences for thermodynamics 
when the requirement of homogeneity is relaxed. In sect. \ref{assum} 
a summary of constructive assumptions is made. 
In sect. \ref{gdsec} 
the generalized Gibbs-Duhem identities are discussed. In sect. 
\ref{ener} the energy representation and the Legendre transform are 
analyzed. In sect. \ref{bhins} some further suggestions 
from black hole thermodynamics are discussed. Sect. \ref{heutd} has 
an heuristic nature, a link between the thermodynamic limit and the 
formalism of quasi-homogeneous thermodynamics is proposed. 
In the Appendix further mathematical and physical aspects 
are explored.

\section{Quasi-homogeneous Pfaffian Forms}
\label{qopf}
  
Given a set of real 
coordinates $x\equiv (x^1,\ldots,x^n)$ and a set of weights 
$\mbox{\boldmath $\alpha$}\equiv (\alpha_1,\ldots,\alpha_n) \in \RR^n$, 
a real-valued function $G(x^1,\ldots,x^n)$ is quasi-homogeneous of degree $r$ 
and type $\mbox{\boldmath $\alpha$}$ if, under the action of 
the one-parameter group of quasi-homogeneous dilatations 
\beq
g^{\tau}: (x^1,\ldots,x^n)\to 
(\hbox{e}^{\alpha_1 \tau}\; x^1,\ldots,\hbox{e}^{\alpha_n \tau}\;x^n),
\eeq 
where $\tau \in \RR$, 
one finds
\beq
G(g^{\tau}\; x)=\hbox{e}^{r \tau}\; G(x).
\eeq
A differentiable function $G$ on a open connected domain 
\beq
{\cal D}=\{  x\in {\cal D}|g^{\tau}  x\in {\cal D}\ \forall\; 
 x\in {\cal D}, \tau\in \RR \}\subseteq \RR^n
\label{dom}
\eeq
is quasi-homogeneous of degree $r$ if and only if \cite{anosov,grudzi}
\beq
D\; G=r\; G,
\label{geul}
\eeq
where the Euler vector field
\beq
D\equiv \alpha_1\; x^1\; \frac{\pa}{\pa x^1}+\ldots+
\alpha_n\; x^n\; \frac{\pa}{\pa x^n}
\label{veuler}
\eeq
is the infinitesimal generator of the transformation. 
Notice that a 
quasi-homogeneous transformation is also called 
``similarity transformation'' and 
``stretching transformation'' \cite{dresner}. 
The identity (\ref{geul}) is a generalization 
of the Euler identity for homogeneous functions. 
Quasi-homogeneity for functions and vector fields is defined 
in Ref. \cite{anosov}. Notice that, according to the definition 
given above, one cannot avoid specifying, together with the degree 
of quasi-homogeneity, the type of the 
quasi-homogeneous object one is considering.\\ 
Homogeneous functions are a subset of 
quasi-homogeneous functions, in fact they have all the weights 
equal to one. We define {\sl strictly quasi-homogeneous functions}  
the functions which satisfy the above definition with weights 
which cannot be all equal.  
A quasi-homogeneous function of $n$ variables is 
characterized by $n$ weights and its degree.  
If all the weights 
are undetermined and arbitrary, or if they are in part 
undetermined and in part equal to one, then all the weights can be 
put equal to one, and then the function is actually homogeneous. 
The viceversa is not true, because there are homogeneous functions which 
don't admit different weights 
(i.e., they are not strictly quasi-homogeneous). A trivial example is 
given by $g(x,y)=x^2+y^2$, which is homogeneous of degree two but 
it is not also strictly quasi-homogeneous. Instead, the function 
$h(x,y)=x y$ is homogeneous of degree two and it is also quasi-homogeneous 
with undetermined weights. In the following, ``quasi-homogeneous'' 
is used as synonymous of strictly quasi-homogeneous, unless a 
more general discussion is required.\\  
A Pfaffian form 
\beq
\omega=\sum_{i=1}^n\; \omega_i(x)\; dx^i
\eeq
is defined to be quasi-homogeneous of degree $r\in \RR$ if, 
under the scaling 
\beq
x^1,\ldots,x^n \mapsto 
\lambda^{\alpha_1}\; x^1,\ldots,\lambda^{\alpha_n}\; x^n
\eeq
one finds
\beq
\omega \mapsto \lambda^r\; \omega.
\eeq
This happens if and only if the degree of quasi-homogeneity 
deg$(\omega_i (x))$ of $\omega_i(x)$ is such that 
deg$(\omega_i (x))=r-\alpha_i\quad \forall i=1,\ldots,n$.  
Let us assume that a quasi-homogeneous $\omega$ is integrable, i.e. $\omega 
\wedge d\omega=0$. 
A quasi-homogeneous transformation is a symmetry for $\omega$ 
(see e.g. \cite{bocharov,cerveau}), in the 
sense that 
\beq
(L_D\; \omega) \wedge\; \omega = 0
\eeq
where $L_D$ is the Lie derivative associated with $D$ defined in 
(\ref{veuler}). 
An integrating factor $f$ such that 
the form $\omega/f$ is closed can then be 
constructed by contracting the vector field $D$ with $\omega$:
\beq
f\equiv \omega(D) =i_D (\omega), 
\eeq 
and $\omega(D)\not \equiv 0$ is to be assumed. This means that 
the symmetry generated by $D$ is nontrivial, in the sense that 
it does not belong to the distribution of codimension one 
which is associated with the kernel of $\omega$. Then, the symmetry 
is not tangent to each leaf of the foliation associated with the 
integrable 1-form $\omega$, but it carries leaves 
onto other leaves. In this sense, a nontrivial symmetry is transversal 
with respect to the foliation.\\  
One gets 
\beq
d\left(\frac{\omega}{f}\right)=0.
\eeq
The integrating factor $f$ is a quasi-homogeneous function of degree 
$r$, because $D\; i_D (\omega) =i_D\; L_D \omega = r\; i_D (\omega)$ 
[due to the Cartan formula $L_X\; i_Y - i_Y\; L_X = i_{[X,Y]}$ and 
to $i_0 =0$]. Then $\omega/f$ is of degree zero and its integral 
can be found only by quadratures (cf. sect. \ref{general}); 
 one gets 
\beq
\has-\has_0\equiv\int_{\Gamma}\; \frac{\omega}{f}.
\eeq
It can be shown that there exists a quasi-homogeneous 
function $F$ of degree one with respect to $D$ such that 
\beqnl
\has&=&\log(F)\\
D\; F&=&F,
\eeqnl 
thus it holds $D\has = 1$ and $\has$ is not quasi-homogeneous. 
On this topic see Appendix A, where the case of a generic transversal 
symmetry is treated. 
The above construction is completely analogous to the construction developed 
for homogeneous integrable Pfaffian forms in Ref. \cite{belhom}. 

\section{a general framework}
\label{general}

Quasi-homogeneous thermodynamics [recall that we mean 
strictly quasi-homogeneous] 
overcomes the standard distinction 
between extensive and intensive variables. We propose the following 
generalization for the integrable Pfaffian form $\deq$ which represents the 
infinitesimal heat exchanged by the system: 
\beq
\deq = dU^{\ast}-\sum_{i=1}^n\; \xi_i^{\ast}\; dX^{i\; \ast}
\label{qoome}
\eeq
where the asterisk indicates that both the independent variables 
$U^{\ast},X^{1\; \ast},\ldots,X^{n\; \ast}$ and the dependent ones 
$\xi_1^{\ast},\ldots,\xi_n^{\ast}$ in (\ref{qoome}) 
are generalizations of the usual ones in the sense that they 
are not simply intensive and/or extensive but quasi-homogeneous, 
in such a way that the Pfaffian form $\deq$ is  
quasi-homogeneous of degree $r$. Moreover, for the sake of 
definiteness and in order to fit some requirements for the validity 
of Frobenius' theorem, $\deq$ is assumed to be of class at least $C^1$ in 
the thermodynamic domain (except, maybe, on the boundary as e.g. the 
surface $T=0$).\\ 
The thermodynamic domain ${\cal D}$ 
is assumed to be a simply connected set which satisfies (\ref{dom}). 
A further requirement for ${\cal D}$ has to be introduced if 
the entropy $S^{\ast}$ is required to be superadditive: ${\cal D}$ 
has to be closed with respect to the sum (see  
sect. \ref{nonh} for further details). Cf. also subsect. \ref{domepi}.  
[In standard homogeneous thermodynamics, 
the thermodynamic domain can be assumed to be a convex cone \cite{belhom}, 
which is a cone with the property to be closed under 
addition (\cite{rockafellar}, pp. 13-14). 
One can also assume that the domain ${\cal D}$ is still a 
convex cone with the further requirement that it 
has to be invariant under quasi-homogeneous transformations. This 
can be obtained by considering a set ${\cal C}\subset \RR^{n+1}$ 
which is invariant under quasi-homogeneous transformations and 
then the set ${\cal K}$ 
of all the positive linear combinations of elements of ${\cal C}$. 
The set ${\cal K}$ is the smallest convex cone containing ${\cal C}$  
(\cite{rockafellar}, p. 14). This cone ${\cal K}$ is trivially 
still invariant under quasi-homogeneous transformations.]\\ 
In the following, it is useful to refer to the 
variables $U^{\ast},X^{1\; \ast},\ldots,X^{n\; \ast}$ as would-be 
extensive variables, 
and to the variables $\xi_1^{\ast},\ldots,\xi_n^{\ast}$ 
(and also $T^{\ast}$) as 
would-be intensive variables. 
The variables 
$U^{\ast},X^{1\; \ast},\ldots,X^{n\; \ast}$ could also be chosen 
to be such that 
they are additive, i.e., if one considers a system which is composed by 
two non-interacting subsystems, then $X^{\ast}=X^{\ast}_1+X^{\ast}_2$.  
See also Ref. \cite{landlett}.\\ 
We assume that the $X^{i\; \ast}$ 
are of degree $\alpha_i$; notice that the degree of $U^{\ast}$ 
is $\alpha=r$, i.e., the degree $r$ of $\deq$ and the weight 
$\alpha$ of $U^{\ast}$ have to coincide in every case. 
The Euler vector field is 
\beq
D= \alpha\; U^{\ast}\; \frac{\pa}{\pa U^{\ast}}
+\sum_i\; \alpha_i\; X^{i\; \ast}\; 
\frac{\pa}{\pa X^{i\; \ast}}.
\label{eulas}
\eeq
An integrating 
factor for $\deq$ is given by 
\beq
f^{\ast}=\alpha\; U^{\ast}-\sum_i\; \alpha_i\; \xi_i^{\ast}\; X^{i\; \ast},
\eeq
and it is assumed that $f^{\ast}\not \equiv 0$ as in standard 
thermodynamics and, moreover, it is assumed that 
$f^{\ast}\geq 0$, which is related to the positivity of the 
absolute temperature. 
Then, it is possible to show that, as in standard thermodynamics, the 
potential $\has^{\ast}$, which is not quasi-homogeneous, is associated 
with a positive definite potential $S^{\ast}$ which is a quasi-homogeneous 
function of degree one with respect to the Euler vector field $D$: 
\beq
\has^{\ast}-\has^{\ast}_0=\int_{\Gamma}\; \frac{\omega}{f^{\ast}}=
\log \left(\frac{S^{\ast}}{S^{\ast}_0}\right),
\label{qos}
\eeq
The proof is found in Appendix A. 

Before continuing our analysis, we recall that a detailed and important 
study on quasi-homogeneous functions and their application 
to scaling and universality is the subject of Ref. \cite{hankey,chang} 
[therein, quasi-homogeneous functions  are called 
``generalized homogeneous functions'' , which is a 
better denomination, but we adopt a common mathematical 
denomination]. 
See also \cite{neff}. 
Further mathematical properties are found in Ref. \cite{aczel} 
[note that, therein, quasi-homogeneous functions are called 
``almost-homogeneous functions'' (\cite{aczel}, p. 231), 
whereas the definition ``generalized homogeneous functions'' 
is assigned to a further generalization of the equation 
defining homogeneous and almost-homogeneous 
functions (\cite{aczel}, p. 304)].

\subsection{metrical entropy}
\label{metrs}

By analogy with the construction for standard thermodynamics, 
a possibility is that $S^{\ast}$ is the metrical entropy for the 
system. This is what happens in 
standard homogeneous thermodynamics and also 
in black hole thermodynamics, 
and we conjecture that this occurrence is not special. See however the 
discussion at the end of this section.\\ 
We define $T^{\ast}\geq 0$ by means of 
\beq
\frac{\pa S^{\ast}}{\pa U^{\ast}}\equiv \frac{1}{T^{\ast}}>0,
\label{tempe}
\eeq
which means that $S^{\ast}$ is assumed to be 
monotonically increasing in $U^{\ast}$. $T^{\ast}$ is another 
integrating factor for (\ref{qoome}), it is such that 
\beq
\deq= T^{\ast}\; dS^{\ast}.
\label{qot}
\eeq
$T^{\ast}$ is a 
quasi-homogeneous function of degree $r-1$ [see also 
theorem 1 of \cite{hankey}, 
where it is shown that the partial derivative $\pa g/\pa X$ 
of a quasi-homogeneous function $g$ of degree $a$ with respect 
to a variable $X$ of weight $b$ is a
quasi-homogeneous function of degree $a-b$ and the same type as $g$]. 
When $r\not =1$, the 
function $T$ is not intensive, but changes under quasi-homogeneous 
rescalings of the system. In the case of a 
Kerr-Newman black hole, one finds e.g. that, 
by doubling the mass and the charge, and by quadruplicating the 
angular momentum, the temperature becomes one half the temperature of the 
original black hole state. This behavior is well-justified in the light 
of the Hawking effect.\\
As a consequence of (\ref{tempe}) and of (\ref{qos}), one finds 
\beq
\frac{\pa \has^{\ast}}{\pa U^{\ast}}=\frac{1}{f^{\ast}}=
\frac{1}{S^{\ast}}\; \frac{\pa S^{\ast}}{\pa U^{\ast}}
=\frac{1}{T^{\ast} S^{\ast}},
\eeq
thus 
\beq
f^{\ast}=T^{\ast} S^{\ast},
\eeq
as in standard thermodynamics. From (\ref{qot}) it is easy to 
show that the quasi-homogeneous entropy can be written as
\beq
S^{\ast}=\alpha\; \frac{U^{\ast}}{T^{\ast}}-\sum_i\; \alpha_i\; 
\frac{\xi^{\ast}}{T^{\ast}}\; X^{i\; \ast}.
\label{saste}
\eeq
In fact, $T^{\ast}$ is an integrating factor of degree $r-1$ for $\deq$, 
and $dS^{\ast}\equiv \deq/T^{\ast}$ is an exact quasi-homogeneous 
Pfaffian form of degree one. Then, as a consequence of lemma 1 
in Appendix \ref{potap} (cf. also result 1 therein), one finds 
$S^{\ast}=i_D\; (\deq/T^{\ast})$, i.e. (\ref{saste}) holds. 
When all the weights in (\ref{saste}) are equal to one, the well-known 
expression for the homogeneous thermodynamic entropy is recovered.

\subsection{thermodynamic foliation} 
\label{folia}
 
As in the case of standard thermodynamics, one can require that 
the thermodynamic foliation is defined by the leaves 
$\has^{\ast}=$ const. everywhere in the thermodynamic manifold, 
except maybe on the boundary. Singularities for $\has^{\ast}$ can 
occur where $f^{\ast}=0$, i.e., in the set 
$Z(f^{\ast})=Z(T^{\ast})\cup Z(S^{\ast})$.\\ 
The surface $T^{\ast}=0$ is expected to represent an adiabatic boundary 
of the thermodynamic manifold. Notice that, as in the 
case of standard thermodynamics, the set $Z(S^{\ast})$ of zeroes of the 
function $S^{\ast}$, if non-empty, is assumed to be contained 
in the set $Z(T^{\ast})$. In standard thermodynamics the occurrence of 
a zero for $S$ at a temperature $T>0$ can be considered pathological, 
a system in such a state should necessarily absorb heat in a neighborhood 
of this state, whichever reversible process could be considered. 
Cf. \cite{belhom}. The same would happen in the quasi-homogeneous case.

Because of (\ref{qos}), the singular values for the thermodynamic 
foliation are represented by the points where $S^{\ast}=0$ (if any).

\subsubsection{a stronger assumption on the domain}
\label{domepi}

One could also assume that 
\beq
f^{\ast}\geq 0 \Longleftrightarrow U^{\ast}\geq 
b(X^{1\; \ast},\ldots,X^{n\; \ast}),
\eeq
where $b(X^{1\; \ast},\ldots,X^{n\; \ast})$ is a quasi-homogeneous 
function of degree $\alpha$ and weights $(\alpha_1,\ldots,\alpha_n)$. 
This function $b(X^{1\; \ast},\ldots,X^{n\; \ast})$ plays the role of 
lowest energy for the system, because by definition $U^{\ast}\geq b$. 
The domain ${\cal D}$ has to include the following set:
\beq
{\mathrm epi}(b)\equiv \{(U^{\ast}, X^{1\; \ast},\ldots,X^{n\; \ast}) 
|  (X^{1\; \ast},\ldots,X^{n\; \ast})\in {\cal K}_{(n)},\quad 
U^{\ast}\geq b(X^{1\; \ast},\ldots,X^{n\; \ast}) \},
\eeq
where ${\cal K}_{(n)}\subseteq \RR^n$ is an open connected set. 
${\mathrm epi}(b)$ is by definition the epigraph of the function $b$. 
If ${\cal K}_{(n)}$ is closed under quasi-homogeneous 
dilatations $(X^{1\; \ast},\ldots,X^{n\; \ast})\mapsto 
(\lambda^{\alpha_1}\; X^{1\; \ast},\ldots,\lambda^{\alpha_n}\; X^{n\; \ast})$, 
then ${\mathrm epi}(b)$ is closed under quasi-homogeneous 
dilatations involving also $U^{\ast}$, because 
$b(\lambda^{\alpha_1}\; X^{1\; \ast},\ldots,
\lambda^{\alpha_n}\; X^{n\; \ast}) = \lambda^{\alpha}\; 
b(X^{1\; \ast},\ldots,X^{n\; \ast})\leq \lambda^{\alpha}\; U^{\ast}$. 
Moreover, if ${\cal K}_{(n)}$ is also closed under the sum and if 
$b$ is subadditive, then  
${\mathrm epi}(b)$ is closed under the sum too, thus the closure 
of ${\cal K}_{(n)}$ under quasi-homogeneous dilatations and under 
the sum allows us to choose ${\cal D}={\mathrm epi}(b)$, in analogy with 
the standard homogeneous case. [Notice that if e.g.
${\cal K}_{(n)}=\RR_+^{n}$, then it is a convex cone which is 
invariant under quasi-homogeneous dilatations]. An example where 
the domain ${\cal D}$ coincides with the epigraph of a quasi-homogeneous 
function is furnished by black hole thermodynamics.\\
Under this assumption on the structure of the domain, one finds that 
it is not possible to find a state having $S^{\ast}=0$ at 
$T^{\ast}>0$, because by introducing  
\beq
B^{\ast}\equiv U^{\ast}-b(X^{1\; \ast},\ldots,X^{n\; \ast})\geq 0
\eeq
one finds 
\beq
S^{\ast}(B^{\ast},X^{1\; \ast},\ldots,X^{n\; \ast})=
S^{\ast}(0,X^{1\; \ast},\ldots,X^{n\; \ast})+
\int_0^{B^{\ast}}\; dY\; 
\frac{1}{T^{\ast}(Y,X^{1\; \ast},\ldots,X^{n\; \ast})},
\eeq
which cannot be zero for any $B^{\ast}>0$ 
(in the last formula mathematical conditions ensuring 
a continuous entropy at $T^{\ast}=0$ 
have been implicitly assumed. Cf. \cite{belhom}).\\

As far as the reference state is concerned, we assume 
that states like the one where 
all the variables  $U^{\ast}, X^{1\; \ast},\ldots,X^{n\; \ast}$ are zero,  
and any states which imply the absence of the system, are unphysical. 
The thermodynamic description starts being meaningful if a statistically 
relevant number of degrees of freedom is available. 
Cf. sect. II of paper \cite{belhom}.

\subsection{possible ambiguities}
\label{ambig}

One may wonder what should happen if 
one finds that $\omega$ is quasi-homogeneous 
but the exact weights are not a priori known. In particular, 
one can assume that their ratio is known, i.e., $\alpha/\alpha_i$ and 
$\alpha_j/\alpha_i$ are known for all $i\not=j$; 
then, there is an overall 
undetermined multiplicative constant $q\not =0$, 
in the sense that these ratios 
don't change if one multiplies all the weights by the same constant. 
One could e.g. know $r^{(1)}\equiv \alpha/\alpha_1$ and 
$r_j^{(1)}\equiv \alpha_j/\alpha_1$ for $\alpha_1\not = 0$ and 
for all $j\not=1$, but the absolute weight $\alpha_1$ is not known, 
thus the weights are $(\alpha_1\; r^{(1)},\alpha_1,\alpha_1\; 
r_2^{(1)},\ldots,\alpha_1\; r_n^{(1)})$.  An example where this 
ambiguity appears is represented by black hole thermodynamics (see 
sect. \ref{bhtd}). One can also assign  
the weights with respect to a scale $\lambda$ whose  
absolute weight could be unknown. 
The absolute weights are then 
of the form $(q\; \alpha, q\; \alpha_1,\ldots,q\; \alpha_n)$ 
with $q$ undetermined. We treat the latter case without loss 
of generality, being the former equivalent to it under the map 
$\alpha_1 \mapsto q\; \alpha_1,\;  q\; \alpha_1\; r^{(1)}
\mapsto q\; \alpha,  q\; \alpha_1\; r_i^{(1)}
\mapsto q\; \alpha_i$ for all $i\not = 1$.  
As a consequence of the undetermined overall factor $q$, 
one has a one-parameter family of 
Euler vector fields $\{D_{(q)} \}$, with 
\beq
D_{(q)}=q\; \alpha\; U^{\ast}\; \frac{\pa}{\pa U^{\ast}}
+\sum_i\; q\; \alpha_i\; X^{i\; \ast}\; \frac{\pa}{\pa X^{i\; \ast}}.
\eeq
Thus, there exist a one-parameter family  $\{f^{\ast}_{(q)} \}$ of integrating 
factors, with $f^{\ast}_{(q)}=\deq (D_{(q)})$, and a one-parameter family 
$\{\has_{(q)} \}$ of potentials such that 
\beq
\frac{\deq}{f^{\ast}_{(q)}}=d \has_{(q)}
\eeq
and a one-parameter family of entropies 
$\{S^{\ast}_{(q)} \}$ satisfying
\beqnl
&& d \has_{(q)}=\frac{dS^{\ast}_{(q)}}{S^{\ast}_{(q)}}\\
&&\cr
&& D_{(q)}\; S^{\ast}_{(q)}=S^{\ast}_{(q)}.
\eeqnl
The true metrical entropy $S^{\ast}$ belongs to the family 
$\{ S^{\ast}_{(q)} \}$ but is undetermined because the overall 
factor $q$ is undetermined. In order to see which relation occurs between 
the various entropies in this family, let us consider a fixed value 
$\bar{q}$ of the parameter $q$. Then we obtain the entropy 
$S^{\ast}_{(\bar{q})}$. If one considers a generic $\hat{q} \not = \bar{q}$, 
one has 
\beq
D_{(\hat{q})}=\frac{\hat{q}}{\bar{q}}\; D_{(\bar{q})} 
\eeq
and 
\beq
f^{\ast}_{(\hat{q})}=\frac{\hat{q}}{\bar{q}}\; f^{\ast}_{(\bar{q})}. 
\eeq
Then, 
\beqnl
&&\frac{\deq}{f^{\ast}_{(\hat{q})}}=\frac{\bar{q}}{\hat{q}}\; 
\frac{\deq}{f^{\ast}_{(\bar{q})}}\\
&&\Longleftrightarrow\cr
&&d\log (S^{\ast}_{(\hat{q})})=\frac{\bar{q}}{\hat{q}}\; 
d\log (S^{\ast}_{(\bar{q})}).
\eeqnl
As a consequence, by assuming in the integral of $\deq/f$ the same 
reference state $(U_0^{\ast},X_0^{1\; \ast},\ldots,X_0^{n\; \ast})$, one finds
\beq
S^{\ast}_{(\hat{q})}(U^{\ast},X^{1\; \ast},\ldots,X^{n\; \ast}) = 
\frac{S^{\ast}_{(\hat{q})}(U_0^{\ast},X_0^{1\; \ast},\ldots,X_0^{n\; \ast})}
{(S^{\ast}_{(\bar{q})}(U_0^{\ast},X_0^{1\; \ast},\ldots,X_0^{n\; \ast}))^{\bar{q}/\hat{q}}}\; 
(S^{\ast}_{(\bar{q})}(U^{\ast},X^{1\; \ast},\ldots,X^{n\; \ast}))^{\bar{q}/\hat{q}},
\eeq
i.e. 
\beq
S^{\ast}_{(\hat{q})}=\zeta_{\hat{q},\bar{q}}\; 
(S^{\ast}_{(\bar{q})})^{\bar{q}/\hat{q}}, 
\label{metrq}
\eeq
where $\zeta_{\hat{q},\bar{q}}\equiv S^{\ast}_{(\hat{q})}
(U_0^{\ast},X_0^{1\; \ast},\ldots,X_0^{n\; \ast})/
(S^{\ast}_{(\bar{q})}(U_0^{\ast},X_0^{1\; \ast},\ldots,X_0^{n\; \ast}))^{\bar{q}/\hat{q}}$ 
is a constant, once one has fixed the reference state. 
The metrical entropy $S^{\ast}$ is related with the entropy 
$S^{\ast}_{(\bar{q})}$ by a simple power law. Cf. also Appendix A.\\
The one-parameter family of temperatures $\{ T^{\ast}_{(q)} \}$, with 
\beq
\frac{1}{T^{\ast}_{(q)}}\equiv \frac{\pa S^{\ast}_{(q)}}{\pa U^{\ast}},
\eeq
is such that 
\beq
\frac{1}{T^{\ast}_{(\hat{q})}}=\zeta_{\hat{q},\bar{q}}\; 
\frac{\bar{q}}{\hat{q}}\;  
(S^{\ast}_{(\bar{q})})^{\bar{q}/\hat{q}-1}\; 
\frac{1}{T^{\ast}_{(\bar{q})}}.
\eeq
Analogous relations exist for the other partial derivatives 
of $S^{\ast}_{(\bar{q})}$ and $S^{\ast}_{(\hat{q})}$. 
Notice that it holds $f^{\ast}_{(q)}=T^{\ast}_{(q)}\; 
S^{\ast}_{(q)}$, because  
\beq
\frac{\pa \has^{\ast}_{(q)}}{\pa U^{\ast}}=\frac{1}{f_{(q)}^{\ast}}=
\frac{1}{S_{(q)}^{\ast}}\; \frac{\pa S_{(q)}^{\ast}}{\pa U^{\ast}}
=\frac{1}{T_{(q)}^{\ast} S_{(q)}^{\ast}}.
\eeq

\subsubsection{phenomenological resolution of the ambiguity}

The metrical entropy $S^{\ast}$ could be phenomenologically 
identified by means of an absolute temperature 
thermometer, made of standard matter, allowing to find, as the 
parameters $U^{\ast}, X^{1\; \ast},\ldots,X^{n\; \ast}$ are varied, 
the function 
$T^{\ast}$ describing the absolute temperature of the 
system. As a consequence, also $S^{\ast}$ can be recovered. 
We remark that the resolution of the above ambiguity by means 
of a phenomenological input is completely in the spirit of 
thermodynamics, which has a phenomenological nature as far as 
statistical mechanics is not taken into account.

\subsubsection{how to fix the absolute weights by means of 
statistical mechanics}
\label{stawei}

A priori, even in statistical mechanics, 
there is a scaling ambiguity, unless the absolute weights are somehow fixed. 
On this topic, see also sect. \ref{heutd}. In fact, let us assume that the 
statistical mechanical entropy [calculated by means of some sort 
of thermodynamic limit or by means of some mean field approximation; 
cf. sect. \ref{heutd}] under rescaling by means of a scale $\lambda$ 
is a quasi-homogeneous function 
of degree $\alpha_S$ and weights $\alpha_1,\ldots,\alpha_n$. 
This means that $S_{sm} (\lambda^{\alpha}\; U,\lambda^{\alpha_1}\; X^1,\ldots,
\lambda^{\alpha_n}\; X^n)=\lambda^{\alpha_S}\; S_{sm} (U,X^1,\ldots,X^n)$. 
By changing the scale $\lambda = 
\bar{\lambda}^q$, the scaling of the variables 
becomes $\bar{\lambda}^{q\; \alpha_i}$. Thus, there is an overall 
factor $q$ of ambiguity. Such a redefinition of the scale, unless 
the absolute weights are somehow fixed, allows to re-map any 
quasi-homogeneous function of degree $\alpha\not =0$ into 
a quasi-homogeneous function of degree one,  
as noticed in \cite{chang}. 
The statistical mechanical entropy is then assumed to satisfy  
\beq
S_{sm} (\lambda^{q\; \alpha}\; U,\lambda^{q\; \alpha_1}\; X^1,\ldots,
\lambda^{q\; \alpha_n}\; X^n)=
\lambda^{q\; \alpha_S}\; S_{sm} (U,X^1,\ldots,X^n).
\label{ssm}
\eeq
where the overall factor ambiguity is enhanced. With respect to 
the Euler vector field $D_{(q)}$ which corresponds to the above weights 
it holds $D_{(q)}\; S_{sm}= q\; \alpha_S\; S_{sm}$.\\ 
We know that from a thermodynamic point of 
view there is a one-parameter family of possible metrical entropies, 
each of which is of degree one. 
The statistical mechanical entropy $S_{sm}$ has to coincide with the 
metrical thermodynamic entropy $S^{\ast}$ and it has to hold 
$T\; dS_{sm}=T^{\ast}\; dS^{\ast}=\deq$. 
As a consequence, the statistical mechanical 
entropy has to be a degree one quasi-homogeneous function. 
This requirement fixes $q\; \alpha_S=1$, 
and fixes unambiguously the weights [then, it eliminates also the 
ambiguity in the one-parameter family of thermodynamic entropies].\\

We could also relax our conjecture about the identification of 
$S^{\ast}$ in formula (\ref{qos}) as the metrical entropy. The 
Clausius-like formula (\ref{qos}) gives in general an empirical 
entropy which is to be uniquely related with the metrical entropy 
by identifying the absolute temperature scale as in the standard 
procedure. One could still assume that the metrical entropy 
is quasi-homogeneous, but it would be possible in line of principle 
to require that its degree is $q$, with $q$ non-necessarily equal to one. 
Then, even in the case where all the weights are known, one would find 
$S^{\ast}_{\mathrm{metric}}\propto (S^{\ast})^q$, where  
$S^{\ast}$ is obtained from formula (\ref{qos}). Cf. also 
Appendix A. The above 
phenomenological procedure could be still allowed. 
Cf. also sect. \ref{heutd} and the conclusions herein.\\

A further consideration is in order, and it concerns the possible 
presence of variables of weight zero. All the mathematical 
approach developed in this section remains unaltered. It is to be 
noted that, if a variable $I^{\ast}$ having weight zero 
is present, the term $I^{\ast}\; \pa/I^{\ast}$ 
does not appear in the Euler operator and in the integrating 
factor $f$ the term proportional to $I^{\ast}$ is missing.\\

Notice that it is possible that the thermodynamic description 
allows a reducing/enlarging of the thermodynamic space 
as the one discussed in Ref. \cite{belhom}. 
This means that 
part of the variables $X^{i\; \ast}$ appearing in $\deq$ and 
in $S^{\ast}$ could be put equal to 
zero consistently, which means that a meaningful thermodynamic 
description is still allowed when $X^{i\; \ast}=0$ 
[in standard thermodynamics such 
variables play e.g. the role of external fields \cite{belhom}]; 
Consistency requires that, if $X^{i\; \ast}=0$ is allowed, 
then $\xi_i^{\ast}=0$ as $X^{i\; \ast}=0$ \cite{belhom}. 
Black hole thermodynamics 
represents a good example, and this topic is discussed in sect. \ref{bhins}.\\

We underline that this kind of generalization 
of thermodynamics  can 
be considered as a sub-case of a generalization proposed 
by P.T.Landsberg in Ref. \cite{landstat}. Therein, a discussion 
of generalized ideal gases appears, where the properties 
of the following generalized entropy are analyzed:
\beq
S=b\; N\; \log(a\; \frac{UV^g}{N^h}). 
\label{lands}
\eeq
Given the Euler field 
\beq
D_{\alpha \beta \gamma}=\alpha U \frac{\pa}{\pa U}+
\beta V \frac{\pa}{\pa V}+\gamma N \frac{\pa}{\pa N},
\eeq
it is easy to show that $S$ in (\ref{lands}) is quasi-homogeneous (of degree 
$\gamma$) if and only if $\alpha+g\; \beta-h\; \gamma=0$ If the latter 
constraint is not implemented, then $S$ is not quasi-homogeneous. 
Homogeneity requires $h=g+1$.

\section{the black hole case}
\label{bhtd}

We summarize herein the results obtained in Ref. \cite{belcar} 
in the black hole case
(we put $\hbar=c=G=k_b=1$, where $k_b$ is the Boltzmann constant; 
moreover, we work with unrationalized electrical units). 
For a general discussion about black hole thermodynamics 
see e.g. \cite{wald,frolov}. Herein, as well as in sect. \ref{hnt} 
and in sect. \ref{geon}, we do not use the asterisk, 
which is in general used only in theoretical sections. 
The Pfaffian form of black hole thermodynamics 
for black holes of the Kerr-Newman family is 
\beq
\deq \equiv dM-\Phi\; dQ-\Omega\; dJ,
\label{dform}
\eeq
where the parameters $M,Q,J$ are the 
mass $M$, the angular momentum $J$ and the 
charge $Q$ of the black hole; the  
angular velocity
\beq
\Omega=\frac{J}{M}\; \frac{1}{2\; M^2-Q^2+2\;M\; \sqrt{M^2-Q^2-J^2/M^2}}  
\eeq
and the electric potential 
\beq
\Phi=\frac{Q\; (M+\sqrt{M^2-Q^2-J^2/M^2})}{2\; M^2-Q^2+2\;M\;  
\sqrt{M^2-Q^2-J^2/M^2}}  
\eeq
are associated with the black hole horizon. 
$M$ in (\ref{dform}) plays the role of internal energy, 
the remaining terms in (\ref{dform}) being ``standard'' work terms. 
Each black hole of the Kerr-Newman family is a stationary solution of the 
Einstein equations, and it can be considered, from the thermodynamic 
point of view we develop here, rather loosely speaking, 
as an ``equilibrium state of the geometry''. The variation occurring in 
(\ref{dform}) is taken along such stationary solutions.  
The thermodynamic domain for the non-extremal black holes is the (open) 
non-extremal manifold implicitly defined by the 
inequality $M^4-M^2\; Q^2-J^2>0$; the extremal 
sub-manifold $M^4-M^2\; Q^2-J^2=0$ 
is a boundary of the former, and its status is analyzed in Ref. 
\cite{belbh} (therein a discussion of the third law of thermodynamics 
also in the light of this approach is found. We limit ourselves to 
recall that in black hole thermodynamics the 
entropic form of the third law is violated \cite{davies} but still the 
unattainability of $T=0$ holds under suitable hypotheses).\\
The Pfaffian form (\ref{dform}) is integrable, 
it satisfies the condition $\deq\; \wedge\; d(\deq)=0$, i.e. 
\beq
-\pa_J\; \Phi+\pa_Q\; \Omega+\Phi\; \pa_M\; \Omega-\Omega\; \pa_M\; \Phi=0.
\label{integrbh}
\eeq
Moreover, the Pfaffian form (\ref{dform}) is 
quasi-homogeneous. In fact, under 
$M\to \lambda^{\alpha} M;\; Q\to \lambda^{\alpha} Q; 
\; J\to \lambda^{2 \alpha} J$, one obtains 
$\deq\to \lambda^{\alpha} \deq$, where $\lambda>0$. 
The weights 
$\alpha,\alpha,2\alpha$ are not known, only their ratio is determined. 
Then, the 
degree of $\deq$ is not fixed \cite{belcar}.  
Nevertheless,   
a thermodynamic construction  is allowed.   
One finds that the integrating factor for $\deq$ is 
given by $f=\alpha\; (M-\Phi\; Q-2\; \Omega\; J)$, that is 
\beq
f=\alpha\; \sqrt{M^2-Q^2-J^2/M^2};
\eeq
[it is interesting to notice that this integrating factor is 
proportional to the horizon coordinate $c$ introduced by B.Carter 
in \cite{cartergeo}. Thus, it is constant on the horizon]. 
Then, one gets \cite{belcar}
\beq
\int\; \frac{\deq}{f}\equiv 
\has-\has_0=\frac{1}{2\alpha}\; \log \left(\frac{A}{A_0} \right),
\label{hasbh}
\eeq
where 
$A=4\; \pi\; (M^2\; (1+\sqrt{1-Q^2/M^2-J^2/M^4})^2+J^2/M^2)$ is the 
black hole area and is positive definite. 
The main 
difficulty is in finding the metrical entropy, because of the  
undetermined overall multiplicative constant $\alpha$. 
%Nevertheless, the additivity of $S$ for black holes in thermal 
%equilibrium is a good tool 
%in order to find the Bekenstein-Hawking law $S=c_0\; A$, which 
%means that $\alpha=1/2$ \cite{belcar}. 
%In fact, the area of two distant black holes 
%is the sum of the areas of each black hole. 
Both $\alpha=1/2$ and 
a undetermined proportionality constant $c_0$ appearing in the  
Bekenstein-Hawking law $S=c_0\; A$ are recovered 
e.g. by comparing the temperature of the black hole with 
the Hawking temperature, and one finds $c_0=1/4$; also the 
Euclidean path integral method can be used with this aim \cite{gihaw}  
(see also the papers collected in \cite{giheu}), 
as well as a phenomenological plot of 
the temperature as a function of $M,Q,J$. 
[Moreover, we notice that 
the same result $S\propto A$ could be obtained by requiring that the 
metrical entropy is quasi-homogeneous of degree one with respect to 
the Euler field $D$ which is obtained by normalizing to one the biggest 
weight $2\alpha$:
\beq
D=\frac{1}{2} M \frac{\pa}{\pa M}+
\frac{1}{2} Q \frac{\pa}{\pa Q}+
J \frac{\pa}{\pa J}.
\eeq
But this normalization rule should be justified on a 
general thermodynamic footing, and the case $J=0$ (Reissner-Nordstr\"{o}m 
solution), being homogeneous in $M,Q$, shows that the physically correct 
value $\alpha=1/2$ cannot be 
found on the ground of such a rule].\\ 
$S$ is quasi-homogeneous of degree one and of type $(1/2,1/2,1)$ and it is 
well defined everywhere in the thermodynamic manifold. It 
is such that $S=0$ only for $M=Q=J=0$, which should not be 
considered as a state belonging to the thermodynamic manifold 
[see the discussion in section \ref{general}; 
moreover, near $M=0$ quantum gravity effects are non-negligible 
and General Relativity is expected to be non-viable]. Thus, 
$Z(S)=\emptyset$ and one finds $Z(f)=Z(T)$, 
and $\has$ is regular everywhere;  as a consequence, 
$\has$ and $S$ could be 
continued even on the extremal submanifold (but see \cite{belbh} for 
arguments against this continuation). 
All non-extremal states have $T>0$, whereas the 
extremal boundary corresponds to $T=0$. As it is well-known, one 
obtains an entropy which is not concave, but simply strictly 
superadditive when the merging of two black holes is 
considered. This ensures the second law of thermodynamics 
\cite{landstat,landtra1,landtra2}. See also sect. \ref{nonh}.\\

By introducing 
\beq
b(Q,J)\equiv \frac{1}{\sqrt{2}}\; \sqrt{Q^2+\sqrt{Q^4+4 J^2}},
\eeq
which is a quasi-homogeneous function of degree $1/2$ and weights $(1/2,1)$, 
as it can be easily verified (see also \cite{belbh}),  
the domain can be also written as follows:
\beq
{\cal D}=\{(M,Q,J)|(Q,J)\in \RR^2,\quad M\geq b(Q,J)\}\equiv {\mathrm epi}(b).
\eeq
In fact, $M=b(Q,J)$ is the physical [i.e. having positive mass] solution  
of $M^4-M^2\; Q^2-J^2=0$ and 
the black hole manifold is $M^4-M^2\; Q^2-J^2\geq 0$, which means 
$M\geq b$. The function $b(Q,J)$ indicates, for given values of $Q,J$, 
which is the lowest mass such that a black hole solution can exist. 
The lowest mass coincides with the mass of the extremal black hole 
having $Q,J$ as charge and angular momentum respectively.\\

We point out that 
the quasi-homogeneous behavior of black hole thermodynamics is 
not a special feature of the Kerr-Newman family. In fact, 
it can be realized easily e.g. in the case of the 
Kerr-Newman-Anti-De-Sitter (KN-AdS) case \cite{caldarelli}. A  
peculiar feature arises, one has to define a 
thermodynamic angular velocity $\Omega$ as the difference between 
the angular velocity of the horizon and the angular velocity 
at infinity, and the same is true in the case of the electric 
potential \cite{caldarelli}. Nevertheless, these definitions 
are necessary, e.g. the ``electric'' work term has to involve 
the difference between the potential at the horizon and the 
potential at infinity when the latter is not zero, and, 
on the same footing, the redefinition of $\Omega$ is due to 
the fact that the angular velocity is not zero at infinity. 
The change in the energy associated with these work terms has 
to involve such a difference, which does not appear in the KN case 
because of the vanishing of $\Phi$ and of $\Omega$ at infinity.\\

General Relativity is surprising from a thermodynamic 
point of view. Black hole solutions of the Einstein 
equations are involved with an integrability condition (\ref{integrbh}),  
and they allow a fine explicit thermodynamic construction \`a la 
Carath\'eodory. One finds a link with thermodynamics which is 
a priori unexpected and it is not simply the formal analogy between 
laws of thermodynamics and laws of black hole mechanics, because 
the above construction belongs to the thermodynamic framework, 
apart from the geometric inputs $\Omega,\Phi$, which come 
from General Relativity, and a comparison with the Hawking effect,   
which fixes $\alpha=1/2$. 
See also the discussion in Ref. \cite{belcar}.

%\newpage

\section{Newtonian gravity. A model of Hertel, Narnhofer and Thirring}
\label{hnt}

There exists an interesting statistical mechanical calculation 
for gravitating fermions which corroborates the idea that, 
in presence of gravity, a quasi-homogeneous thermodynamics is 
allowed \cite{hertel,hert1,thirbook}. The model (HNT model in the following) 
involves non-relativistic 
fermions interacting by means of a Coulomb potential and a 
Newtonian potential, with vanishing [or small] 
total charge. Thus, the calculation does not involve General 
Relativity but only Newtonian gravity. Nevertheless, the 
result is still interesting and, to some extent, puzzling. 
The quantum mechanical Hamiltonian which is considered is 
\beq
H=\sum_{i=1}^{N}\; \frac{\vec{p}_i^2}{2 M_i}+ 
\sum_{i<j}\; \frac{e_i e_j-G m_i m_j}{|\vec{x}_i-\vec{x}_j|}.
\label{hamhe}
\eeq
The authors state that a thermodynamic limit exists 
if the following asymptotic behavior is allowed for $U,V$: 
$V\sim 1/N$, $U\sim N^{7/3}$, and the 
micro-canonical entropy is of degree one in 
the number $N$ of fermions. 
The scaling properties of $U$, $V$ with 
$N$ are peculiar, they are homogeneous of degree one if the potential 
term in (\ref{hamhe}) 
is suppressed by means of a multiplicative factor $1/N^{2/3}$, which means 
that, as the system gets larger, the interaction becomes weaker 
\cite{thirbook}. 
If this suppression is not in agreement with the physics, and if the 
fermions do not become relativistic, then the micro-canonical 
entropy $S(U,V,N)$ 
is a quasi-homogeneous function of degree one, 
and the weights of 
$U,V,N$ are $7/3,-1,1$ respectively. Cf. also sect. \ref{heutd}, 
where the notion of quasi-homogeneous thermodynamic limit is introduced. 
The Euler field is
\beq
D_{UVN}=\frac{7}{3}\; U\; \frac{\pa}{\pa U}-
V\; \frac{\pa}{\pa V}+N\; \frac{\pa}{\pa N}.
\eeq
For details of the highly non-trivial 
calculations, see Refs. \cite{hertel,hert1,messer}. We are interested here 
in pointing out that, as stressed in Ref. \cite{hertel}, there is 
a one-parameter family of equivalent limits $N\to \infty$, whose 
parameter $\gamma$ remains fixed by the requirement that the 
ground state energy goes like $N^{7/3}$, 
according to a result of Ref. \cite{leblond}. A scaling 
ambiguity, which would reflect itself in the undetermined weights 
$(-2\gamma +5/3,3\gamma,1)$ of $(U,V,N)$ respectively, 
is thus solved by comparing with 
the scaling behavior of the ground state. See also \cite{messer}. 
Another important point
is that the temperature-dependent Thomas-Fermi equation, 
which allows these scaling 
properties, becomes exact in the thermodynamic limit for the 
system under discussion \cite{hertel}.\\   
We limit ourselves to underline that 
Newtonian gravity furnishes a statistical mechanical 
example where unusual scaling laws for the ``extensive'' variables 
are allowed. The purely attractive nature of gravity plays a 
major role, because it does not saturate, i.e., it does not 
allow to obtain a ground state proportional to the number of 
particles \cite{leblond}.  
Notice also that a lack of concavity is allowed \cite{thirbook}.\\ 
It is interesting to notice that another model involving 
different scaling relations has been developed in Ref. \cite{vega}. 
Therein, a classical gas of non-relativistic particles which 
interact by means of Newtonian gravity is considered in a 
diluted regime where particles of mass $m$ are enclosed in a 
box; the following behavior is recovered in a non-conventional 
thermodynamic limit where $V/N^3 \to $const. as $N\to \infty$. 
A dilution parameter 
\beq
\xi \equiv \frac{1}{G\; m^2}\; \frac{U\; V^{1/3}}{ N^2}
\label{dilu}
\eeq
is kept constant in the thermodynamic limit. 
One finds that $S(U,V,N)$ is quasi-homogeneous of degree one and 
weights $(1,3,1)$, as can be inferred from Ref. \cite{vega}. Cf. also 
sect. \ref{heutd}. 
This model differs from the previous one 
because it involves classical matter, the gas is kept diluted in the 
thermodynamic limit, no collapse is considered. 
Note that even the HNT model can be related with the parameter $\xi$ 
introduced in (\ref{dilu}), in fact even in HNT model this parameter 
is constant in the appropriate thermodynamic limit. 
$\xi$ can be qualitatively interpreted as  
the ratio between 
the internal energy and the Newtonian gravitational energy, in fact the 
gravitational energy of a homogeneous sphere of mass $M\equiv N^2\; m$ 
is proportional to $G M^2/V^{1/3}$. Thus, in both cases, the 
requirement of a non-negligible contribution of the gravitational 
energy to the internal energy is realized in the thermodynamic limit. 
In the HNT model, a self-bounded system is considered, like a (cold) star, 
where $V$ contracts as the mass is increased; in the model of Ref. 
\cite{vega}, the physical conditions are different, a gaseous system 
is considered, the energy per particle  
remains constant and small, and the systems remains gaseous in the 
thermodynamic limit.\\ 
In the regime where gravitational forces cannot be neglected, 
one can expect that a different kind of quasi-homogeneous thermodynamics 
can arise, with different weights, depending on the actual physical 
conditions, e.g., the amount and the type 
of matter. Semi-relativistic matter is expected to behave in a 
different manner (cf. Ref. \cite{leblond}). 
The point is that a different scaling behavior 
of the ground-state of the matter can be expected if relativistic 
effects in the kinetic energy are non-negligible. 
A self-consistent calculation 
taking into account General Relativity is expected to give 
another kind of contribution. 
An approach, combining in a self-consistent way general-relativistic 
equations and finite-temperature Thomas-Fermi approximation for 
the thermodynamic functionals would be required. In this direction, 
see Ref. \cite{bilic}. It is 
also interesting to point out that such an approach probably would not 
be useful for understanding black hole thermodynamics, in fact no 
stable ground state for the matter could exist by hypothesis 
(a collapse of the matter beyond the possible stable states represented 
by white dwarf and neutron stars should take place, with no possibility 
to prevent the formation of an event horizon).

\section{a look at the thermal geon and at self-gravitating radiation}
\label{geon}

We discuss naively the scaling properties of the so-called thermal geons 
\cite{power}. A geon is a ``gravitational electromagnetic entity'' which 
consists in a self-gravitating ensemble of electromagnetic modes 
\cite{wheeler}. 
The most stable 
configuration is such that the photons are distributed within a toroidal 
region of the space-time, but also a spherical distribution can be allowed. 
Geons don't represent strictly stationary solutions of the Einstein equations, 
because a leakage of photons to infinity is allowed. Nevertheless, 
they can be considered as metastable quasi-stationary solutions. 
Moreover, sizes are considered where no contribution of electron 
physics, due to pair creation by the electromagnetic field, is considered, 
and no zero-point energy is taken into account. 
In Ref. \cite{power} the idea of a geon has been generalized to the 
case of thermally distributed electromagnetic modes. The main assumptions 
are the following ones: a) the metric is a spherosymmetric static diagonal 
metric in the Schwarzschild gauge; b) there exist two separate classes of 
electromagnetic modes; the first class is constituted by bounded null 
geodesics which represent modes whose energy cannot escape to infinity 
[the actual rate of flux to infinity goes to zero exponentially  with the 
ratio between the dimensions of the geon and the wavelength $\lambda$]. 
The second class is constituted by free electromagnetic modes, i.e. by 
null geodesics which reach infinity. c) The energy of the free modes is 
zero, the energy of the bounded modes of frequency $\Omega$ 
is [in natural units]
\beq
E_{\Omega}=\Omega\; \frac{1}{\exp \left(\frac{\Omega}{T} \right)-1},
\eeq
which is the usual distribution for black body radiation and $T$ is 
the temperature. See also \cite{power} on this topic. 
The active region of the geon is defined as the region 
where the metric coefficient $g_{00}$ satisfies $|g_{00}|<1$. 
By defining $R$ as the radius of the active region 
and $M$ as its mass, 
the following scaling laws can be deduced \cite{power}:
\beqnl
R\sim R_T&\equiv & \left(\frac{15 \hbar^3 c^7}{8 \pi^3 G} \right)^{1/2}\; 
\frac{1}{T^2}
\propto \frac{1}{T^2}\\
M\sim M_T&\equiv & \left(\frac{15 \hbar^3 c^{11}}{8 \pi^3 G^3} \right)^{1/2}\; 
\frac{1}{T^2} \propto \frac{1}{T^2}.
\eeqnl
As an order of magnitude estimation on the same footing of the previous 
evaluations, one can see that the entropy behaves as $S\propto V T^3$, 
where $V$ is the volume of the active region, thus $S\propto T^{-3}$. 
From the above naive considerations, one finds that $S$ is a quasi-homogeneous 
function
\beq
S(\lambda^{-2}\; M, \lambda^{-6}\; V)=\lambda^{-3}\; S(M,V);
\eeq
In the case of a massless gas, there is an ambiguity in the 
identification of the absolute weights of the variables. 
There is an overall multiplicative factor to be determined in 
the Euler vector field.   
Let us assume that 
$S$ is of degree one, as suggested by the formal 
picture of thermodynamics (cf. sect. \ref{general} and sect. \ref{heutd}); 
then $T$ has degree $-1/3$ 
and the scaling law for the fundamental equation is 
\beq
S(\lambda^{2/3}\; M, \lambda^2\; V) = \lambda\;  S(M,V),
\eeq
i.e., $M$ has weight $2/3$ and $V$ has weight $2$. We note that the ratio 
$M/V^{1/3}$ is scale-invariant and it is the same ratio which is kept 
constant in the thermodynamic limit discussed in \cite{vega}.\\
The above result about thermal radiation can be confirmed by comparing it 
with the study of Ref. \cite{sorkw}. Black body self-gravitating radiation 
enclosed in a spherical box of radius $R$ is considered in \cite{sorkw}. 
It is shown that the maximization of the entropy in a spherosymmetric 
geometry leads to the following results: i) the metric is static; 
ii) the perfect fluid of photons satisfies the Tolman-Oppenheimer-Volkov 
(TOV) equation. In particular, 
the TOV equation one obtains is scale-invariant. 
By introducing the density $\rho(r)$, the mass 
$m(r)=\int_0^r dy 4 \pi y^2 \rho(y)$ ($r$ is the radial coordinate) one can 
define, by following \cite{sorkw}, the dimensionless variables 
\beqnl
\mu &=& m(r)/r;\\
q &=& dm/dr = 4\pi r^2\; \rho(r);
\eeqnl 
and the TOV becomes equivalent to the following couple of 
equations:
\beqnl
r \frac{dq}{dr} &=& \frac{2 q}{1-2\mu}\; (1-4\mu -\frac{2}{3}\; q),\\
r \frac{d \mu}{dr} &=&q-\mu.
\label{tovrad}
\eeqnl
No equilibrium is attained if $\mu>0.25$ \cite{sorkw}.  
We are interested here in the scaling properties of the Pfaffian form 
\beq
\deq = dM + p\; dV,
\eeq
where $M\equiv m(R)$ is the mass, $V$ is the volume 
(as seen from infinity) and $p=\rho(R)/3$ is the pressure as a function of 
the box radius. These variations are to be intended, as in the black hole 
case, as ``on shell'' , i.e., along static spherosymmetric equilibrium 
solutions of the TOV. Each solution represents a thermodynamic equilibrium 
state. 
Under the scaling 
$M \to \lambda\; M;\; V \to \lambda^3\; V;\; p \to \lambda^{-2}\; p$, 
which corresponds to (30) of \cite{sorkw} and  
is allowed by the scale invariance of 
the TOV, one finds $\deq \to \lambda\; \deq$. Particularly, one has 
\beq
D=\alpha\; (M\; \frac{\pa}{\pa M}+ 3\; V\; \frac{\pa}{\pa V}),
\eeq
where $\alpha$ has to be determined. From $dV = 4\pi R^2\; dR$ 
and from  (\ref{tovrad}) one finds 
\beq
\deq = \frac{4}{3}\; q\; dR
\eeq
and the integrating factor
\beq
\deq(D)=\alpha\; (M+3 p V) = \alpha\; \frac{1}{3}\; (3\mu+q)\; R.
\eeq
Thus, 
\beq
\frac{\omega}{f}=\frac{4}{\alpha}\; \frac{q}{3\mu+q}\; \frac{dR}{R}. 
\eeq
By comparing this result with the ratio $dS/S$ which can be obtained by 
the exact result (34) of \cite{sorkw}, one finds that $\alpha = 2/3$. 
As the consequence, the weights of $D$ are the same as in the case of 
thermal geons (notice that the requirement for $S$ to be quasi-homogeneous 
of degree one in the scale factor $\lambda$ would lead to the scaling 
$r \to \lambda^{2/3}\; r; m \to \lambda^{2/3}\; m;\; 
\rho \to \lambda^{-4/3}\; \rho$ in (30) of \cite{sorkw}).

%\newpage

\section{consequences of the non-extensivity of $S$}
\label{nonh}

We have to underline that the lack of extensivity 
has important consequences on the properties of the entropy. In fact, 
if one considers for a continuous entropy $S$ 
superadditivity (S), concavity (C) and homogeneity (H), one finds:\\
\\ 
a) (H) and (S) imply (C) \cite{galg};\\ 
b) (H) and (C) imply (S)\cite{land};\\ 
c) (S), with the condition $S(0)=0$, and (C) imply (H) 
\cite{landlett,landtra1}.\\ 
\\ 
\noindent
We recall that a function $h(x)$ 
of $n$ variables, collectively indicated with 
$x$ in this section, is superadditive if 
$h(x+y)\geq h(x)+h(y)$ for all $x,y$ in its domain. Superadditivity, 
in the case of entropy, means that the principle of entropy increase 
in a irreversible adiabatic process %for the system 
is respected. 
Moreover, (C) requires a convex domain, (S) a domain which has to be 
closed under addition, and (H) a domain which has to be closed under 
multiplication of $x$ by a real scalar $\lambda>0$ (a cone).\\ 
\\
Notice that the condition $S(0)=0$ appearing in c) is a mathematical condition 
which could be required even if the state $x=0$ does not belong 
to the thermodynamic space, in the light of our discussion in  
sect. \ref{general}; see however also \cite{landtra1}, where $x=0$ 
is included in the thermodynamic space. One could consider an 
extended domain including $x=0$, on a purely mathematical ground, 
even if the thermodynamic formalism at such a point is meaningless. 
Moreover, $S$ is required to be a continuous function. 
Then, if homogeneity does not hold and $S(0)=0$, 
either superadditivity or concavity has to be violated. It can be 
easily shown that the condition $S(0)=0$ in c) can be replaced by the 
(natural) requirement that $S(x)\geq 0$ for any state $x$ in the 
thermodynamic domain. In fact, (C) implies 
\beq
S(\frac{1}{2}\; y + \frac{1}{2}\; z)\geq  
\frac{1}{2}\; S(y)+\frac{1}{2}\; S(z)
\label{midconv}
\eeq
for any $y,z$; moreover, (S) implies 
\beq
S (2\; x) \geq 2\; S(x).
\eeq
If $z=0$ and $y=2\; x$, from (\ref{midconv}) one obtains
\beq
S(x)\geq  \frac{1}{2}\; S(2\; x)+\frac{1}{2}\; S(0).
\eeq
Then, $1/2\; S(2\; x)+1/2\; S(0)\leq S(x) \leq 1/2\; S(2\; x)$, 
which implies $S(0)\leq 0$. As a consequence, the requirement of a 
nonnegative entropy $S$, together with (S) and (C), implies 
$S(0)=0$, thus also (H) has to hold. This shows that the condition 
$S(0)=0$, in the framework of thermodynamics, where $S\geq 0$, 
cannot be actually considered as a true restriction leaving room 
for a thermodynamics in which a concave and superadditive but 
non-homogeneous entropy is allowed if $S$ is defined in $x=0$ 
[of course, this holds as far as 
negative values of $S$ are forbidden].  
Notice that $S^{\ast}(0)=0$ for a continuous quasi-homogeneous 
entropy defined in $0$.\\

Superadditivity 
of the entropy means that for the thermodynamic system the entropy 
does not increase by fragmenting the system \cite{landlett}. 
From the point of view 
of the energy representation, to be discussed in sect. \ref{ener}, 
superadditivity of the entropy is equivalent to the subadditivity of 
the energy under the condition (\ref{tempe}) [for a proof, see 
Refs. \cite{galg,land}. This equivalence holds even if conditions 
(C), (H) are violated]. 
Subadditivity of the energy means that exploding is not 
energetically advantageous for the system \cite{thirpre,thirna}.  
The relevance of (S) for the second law of thermodynamics 
is discussed and underlined in Ref. \cite{landstat}. 
Accordingly, 
one should privilege the superadditivity property against the concavity, 
and superadditivity should be required as a fundamental 
property for the quasi-homogeneous picture of thermodynamics.  
Notice that a lack of concavity for thermodynamics in presence of 
gravity is verified in Refs. \cite{lynden,thirring}, where 
negative heat capacities in presence of gravity are calculated.  
Therein, a discussion on how to deal with the lack of the 
standard stability properties of thermodynamics is sketched. 
See also \cite{padma,chavanis}. It is to be noted that this lack of 
concavity in presence of gravity forces to abandon the 
usual Gibbsian scheme for 
thermodynamics, and the homogeneity property has to be withdrawn 
because of a), if one adopts the superadditivity as a fundamental 
property for the metrical entropy of a system. Due to the above 
equivalence between superadditivity of the entropy and subadditivity 
of the internal energy, one can naively justify this choice by  
underlining that, because of the purely attractive nature of gravity, 
internal energy of a self-gravitating system should be strictly subadditive, 
in fact energy decreases as the accretion of matter increases. 
Thus, there is an implosive instability of gravity, which eventually 
leads to the formation of black holes, to be identified as very long-lived 
metastable states. By passing, we note that there is another instability 
which is opposite in character with respect to the one implied by 
gravity. It involves system fragmentation/explosion because of a 
superadditive energy \cite{thirpre,thirna}. 
It should be characterized only by unstable states 
[systems with a suitable charge excess would be explosive \cite{messer}. 
See also the works of E.H.Lieb collected in Ref. \cite{liebook}, in 
particular the review \cite{liermp}; see also \cite{lieleb}]. 

%\vskip -0.5truecm
%\newpage

\section{summary of constructive assumptions}
\label{assum}

We summarize the set of constructive assumptions upon which 
our approach is based. They correspond to the assumptions leading 
to standard homogeneous thermodynamics \cite{belhom}, the difference consists 
in the substitution of the homogeneity symmetry with the quasi-homogeneity 
symmetry and also in the explicit request that the entropy is superadditive. 
We comment only some assumptions which require further remarks with 
respect to the discussion developed in the text.\\ 
\\
{\bf a1}) 
{\sf The quasi-homogeneity symmetry which characterizes the thermodynamic 
system is translated into the quasi-homogeneity of the Pfaffian form} 
\quad $\deq = dU^{\ast}-\sum_{i=1}^n\; \xi_i^{\ast}\; dX^{i\; \ast}$.\\
\\ 
{\bf a2}) {\sf The quasi-homogeneity symmetry of $\deq$ is 
nontrivial}.\\ 
\\ 
{\bf a3}) {\sf  
The thermodynamic foliation is defined by the leaves $\has^{\ast\; -1}(c)$, 
where $c\in \RR$ is a constant,   
everywhere in ${\cal D}$.}\\ 
\\ 
This assumption does not leave room for foliations which are 
based on a quasi-homogeneous $S^{\ast}$ which is allowed to be negative. 
Cf. the discussion in \cite{belhom} for the homogeneous case.\\
\\
{\bf a4}) {\sf The integrating factor $f^{\ast}$ is 
non-negative}.\\ 
\\ 
{\bf a5}) {\sf The metrical entropy $S^{\ast}$ is 
quasi-homogeneous of degree one}.\\ 
\\
This assumption could also be weakened, as discussed in the text.\\
\\ 
{\bf a6}) {\sf The metrical entropy $S^{\ast}$ is 
superadditive}.\\ 
\\
This assumption substitutes the requirement for a concave entropy 
occurring in standard homogeneous thermodynamics. Notice that 
in standard thermodynamics it would lead again to a concave metrical 
entropy, because of the implication a) of sect. \ref{nonh}. Cf. \cite{galg}.\\ 
\\
{\bf a7}) 
{\sf We require that $Z(S^{\ast})\subseteq Z(T^{\ast})$.}\\ 
\\ 
This assumption is e.g. implemented as in sect. \ref{domepi}.\\
One can also add the following assumptions which appear to be 
physically appealing:\\ 
\\
{\bf a8}) 
{\sf We require that to each level set $S^{\ast}=$ const. corresponds a unique 
leaf.}\\
\\
This assumption means that two states of the same quasi-homogeneous 
system lying on the same isoentropic surface are path-connected, i.e. 
it is possible to find a reversible adiabatic path connecting each other.\\
\\ 
{\bf a9}) 
{\sf We require that $\frac{\pa S^{\ast}}{\pa V^{\ast}}$ is positive if 
$V^{\ast}$ is the volume of the system.}\\
\\  
This simply means that the pressure $p^{\ast}$ is positive whenever 
it is definite. \\
\\ 
{\bf a10}) 
{\sf We require that $f^{\ast}=0$ corresponds to an integral 
manifold of $\deq$.}\\
\\  
This assumption is natural and is related to the problem of the 
third law. Cf. \cite{belhom} for standard thermodynamics; see also 
\cite{belbh} for the black hole case and also 
Appendix \ref{thirdap}.

\newpage

\section{generalized Gibbs-Duhem equation}
\label{gdsec}

Quasi-homogeneity of $S^{\ast}$ allows to find  a generalization 
of the standard Gibbs-Duhem equation \cite{callen}. 

\subsection{case where none independent variable of weight zero appears}

We first discuss the case 
where none independent variable of weight zero appears. 
By differentiating (\ref{saste}) one finds 
\beq
dS^{\ast}=\alpha\; \frac{1}{T^{\ast}}\; dU^{\ast}+\alpha\; 
U^{\ast}\; d\left(\frac{1}{T^{\ast}}\right)-
\sum_i\; \alpha_i\; \frac{\xi_i^{\ast}}{T^{\ast}}\; dX^{i\; \ast}
-\sum_i\; \alpha_i\; X^{i\; \ast}\; 
d\left(\frac{\xi_i^{\ast}}{T^{\ast}}\right) 
\eeq
and by comparing with 
\beq
dS^{\ast}=\frac{1}{T^{\ast}}\; dU^{\ast}-\sum_i\;  
\frac{\xi_i^{\ast}}{T^{\ast}}\; dX^{i\; \ast}
\label{diffes}
\eeq
one finds that the following generalized Gibbs-Duhem equation
has to hold:
\beq
\alpha\; {U^{\ast}}^{-\frac{\alpha-1}{\alpha}+1}\; 
d\left(\frac{{U^{\ast}}^{\frac{\alpha-1}{\alpha}}}{T^{\ast}}\right) 
-\sum_i\; \alpha_i\; 
{X^{i\; \ast}}^{-\frac{\alpha_i-1}{\alpha_i}+1}\;
d\left(\frac{{X^{i\; \ast}}^{\frac{\alpha_i-1}{\alpha_i}}\; \xi_i^{\ast}}{T^{\ast}}\right)=0. 
\label{gdg}
\eeq
This equation, as in the usual thermodynamic case, allows to 
express the differential of a would-be intensive 
variable as a function of all the other ones, and then to find 
such a would-be intensive variable apart from a integration constant 
(cf. \cite{callen} for the case of standard thermodynamics). 
In fact, one could e.g. find for $1/T^{\ast}$:
\beq
d\log \left(\frac{1}{T^{\ast}}\right)=-\frac{\alpha-1}{f^{\ast}}\; dU^{\ast}
+\frac{1}{f^{\ast}}\; \sum_i\; \alpha_i\;
{X^{i\; \ast}}^{-\frac{\alpha_i-1}{\alpha_i}+1}\;
d\left({X^{i\; \ast}}^{\frac{\alpha_i-1}{\alpha_i}}\; \xi_i^{\ast}
\right).
\label{ggd}
\eeq
Notice that 
\beq
d\log \left(\frac{1}{T^{\ast}}\right)=-d\log (f^{\ast})+\frac{\deq}{f^{\ast}}.
\label{ggdint}
\eeq
Eqn. (\ref{ggd}) is implemented if (\ref{qos}) and (\ref{diffes}) hold.  
Let us consider the inverse problem where one assigns $n$ would-be intensive 
functions $\xi_1^{\ast},\ldots,\xi_n^{\ast}$ which are quasi-homogeneous 
in $U^{\ast},X^{1\; \ast},\ldots,X^{n\; \ast}$ and such that 
$\deq$ is quasi-homogeneous as well. Eqn. (\ref{ggd}), or, 
equivalently, eqn. (\ref{ggdint}) can be used for recovering 
$T^{\ast}$, and then for reconstructing $S^{\ast}$ by means of (\ref{qos}), 
if and only if one has ensured that 
$\deq/f^{\ast}$ is an exact differential, i.e., that $\deq$ is 
integrable and $f^{\ast}$ is an integrating factor for $\deq$. 
Otherwise, (\ref{ggd}) is not even meaningful (a closed form 
on the left side of (\ref{ggdint}) should be equal to a non-closed form 
on the right side of the same equation). 
Cf. also \cite{belhom}. 
\\
Notice that, if $S^{\ast}$ has degree $q$, then (\ref{gdg}) becomes
\beq
\frac{\alpha}{q}\; {U^{\ast}}^{-\frac{\alpha-q}{\alpha}+1}\; 
d\left(\frac{{U^{\ast}}^{\frac{\alpha-q}{\alpha}}}{T^{\ast}}\right) 
-\sum_i\; \frac{\alpha_i}{q}\; 
{X^{i\; \ast}}^{-\frac{\alpha_i-q}{\alpha_i}+1}\;
d\left(\frac{{X^{i\; \ast}}^{\frac{\alpha_i-q}{\alpha_i}}\; \xi_i^{\ast}}{T^{\ast}}\right)=0. 
\label{gdes}
\eeq
See also Appendix \ref{gdap}.

\subsection{case where independent variables of weight zero appear}

Let us assume that,  the 
independent variables $X^{i\; \ast}$ 
have weight zero for $i=p+1,\ldots,n$, with $p<n$. This means that 
$\xi_i^{\ast}$ have degree $\alpha$ for $i=p+1,\ldots,n$. 
The Euler vector field is 
\beq
D=\alpha\; U^{\ast}\frac{\pa}{\pa U^{\ast}}+\sum_{i\leq p}\; \alpha_i\; 
X^{i\; \ast} \frac{\pa}{\pa X^{i\; \ast}},
\eeq
and the integrating factor is 
\beq
f^{\ast}=
\alpha\; U^{\ast}- \sum_{i\leq p}\; \alpha_i\; \xi_i^{\ast}\; X^{i\; \ast}.
\eeq
Then, one has 
\beq
S^{\ast}=\frac{1}{T^{\ast}}\; (\alpha\; U^{\ast}-
\sum_{i\leq p}\; \alpha_i\; \xi_i^{\ast}\; X^{i\; \ast}).
\eeq
It is easy to show that the Gibbs-Duhem equation in this case is 
\beq
\alpha\; {U^{\ast}}^{-\frac{\alpha-1}{\alpha}+1}\; 
d\left(\frac{{U^{\ast}}^{\frac{\alpha-1}{\alpha}}}{T^{\ast}}\right) 
-\sum_{i\leq p}\; \alpha_i\; 
{X^{i\; \ast}}^{-\frac{\alpha_i-1}{\alpha_i}+1}\;
d\left(\frac{{X^{i\; \ast}}^{\frac{\alpha_i-1}{\alpha_i}}\; \xi_i^{\ast}}{T^{\ast}}\right)+ \sum_{p<i\leq n}\; 
\frac{\xi_i^{\ast}}{T^{\ast}}\; dX^{i\; \ast}=0. 
\label{gdgw}
\eeq
One can also obtain 
the analogous of eqn. (\ref{ggd})
%\newpage
\beqnl
d\log \left(\frac{1}{T^{\ast}}\right)&=&-\frac{\alpha-1}{f^{\ast}}\; 
dU^{\ast}
+\frac{1}{f^{\ast}}\; \sum_{i\leq p}\; \alpha_i\;
{X^{i\; \ast}}^{-\frac{\alpha_i-1}{\alpha_i}+1}\;
d\left({X^{i\; \ast}}^{\frac{\alpha_i-1}{\alpha_i}}\; \xi_i^{\ast}
\right)\cr
&-&\frac{1}{f^{\ast}}\; \sum_{p<i\leq n}\; \xi_i^{\ast}\; dX^{i\; \ast}.
\label{giduw}
\eeqnl
(\ref{ggdint}) still holds. 

Concerning the Gibbs-Duhem equation in standard thermodynamics, 
see \cite{qotn}.

\subsection{examples}

In the black hole case, as well known, one has [we don't write  
the asterisk in what follows] 
\beq
S=\frac{M}{2 T}-\frac{\Phi Q}{2 T}-\frac{\Omega J}{T}
\eeq
and 
\beq
dS=\frac{1}{T}\; dM-\frac{\Phi}{T}\; dQ-\frac{\Omega}{T}\; dJ; 
\eeq
the generalized Gibbs-Duhem equation, associated with the 
quasi-homogeneity of $S$, is 
\beq
-\frac{1}{2 T}\; dM + \frac{M}{2}\; d\left( \frac{1}{T}\right)
+\frac{\Phi}{2 T}\; dQ 
-\frac{Q}{2}\;  d\left(\frac{\Phi}{T}\right) 
- J\; d\left(\frac{\Omega}{T}\right)=0,
\eeq
which can be rewritten as \cite{belcar}
\beq
\frac{1}{2}\; M^2\; d\left( \frac{1}{M T}\right)
- \frac{1}{2}\; Q^2\; d\left(\frac{\Phi}{Q T}\right) 
- J\; d\left(\frac{\Omega}{T}\right)=0.
\eeq
Then, one can find $1/T$ from 
\beq
d\log \left(\frac{1}{T} \right)=\frac{1}{f}\left( \frac{1}{2}\; dM+
\frac{1}{2}\; Q^2\; d \left(\frac{\Phi}{Q} \right) + J\; d\Omega \right).  
\eeq
It can be easily shown that $1/T$ can be determined apart from an 
undetermined multiplicative constant.

\section{Energy representation and Legendre transform}
\label{ener}

In the energy representation, the fundamental equation is 
$U^{\ast}=U^{\ast}(S^{\ast},X^{1\; \ast},\ldots,X^{n\; \ast})$. One has 
\beq
dU^{\ast}=T^{\ast}dS^{\ast}+\sum_i\; \xi_i^{\ast}\; dX^{i\; \ast}.
\label{qoexa}
\eeq
$U^{\ast}$ is a quasi-homogeneous function of degree $r$ 
and type $(1,\alpha_1,\ldots,\alpha_n)$; $T^{\ast}$ is quasi-homogeneous 
of degree $r-1$, $\xi_i^{\ast}$ are quasi-homogeneous of degree 
$\alpha_i - r$. (\ref{qoexa}) represents an exact quasi-homogeneous 
Pfaffian form of degree $r$. The Euler operator is 
\beq
D = S^{\ast}\; \frac{\pa }{\pa S^{\ast}}+\sum_i\; \alpha_i\;
X^{i\; \ast}\; \frac{\pa }{\pa X^{i\; \ast}}.
\eeq
One has then
\beq
U^{\ast}=\frac{1}{r}\; 
(T^{\ast} S^{\ast}+\sum_i\; \alpha_i\; \xi_i^{\ast}\; X^{i\; \ast}).
\eeq
[Notice that the generalization to the case where $S^{\ast}$ is 
quasi-homogeneous of degree $q$ is trivial].
\\
As far as the Gibbs-Duhem equation is concerned, one easily finds 
[we shift to the case where $S^{\ast}$ has degree $q$]
\beq
\frac{q}{r}\;  {S^{\ast}}^{-\frac{q-r}{q}+1}\; 
d\left({S^{\ast}}^{\frac{q-r}{q}} T^{\ast}\right)
+\sum_i\; \frac{\alpha_i}{r}\; 
{X^{i\; \ast}}^{-\frac{\alpha_i-r}{\alpha_i}+1}\;
d\left({X^{i\; \ast}}^{\frac{\alpha_i-r}{\alpha_i}}\; \xi_i^{\ast}\right)=0. 
\label{gdee}
\eeq
For a proof see Appendix \ref{gdap}. 
One can e.g. determine $T^{\ast}$ from 
\beq
d\left({S^{\ast}}^{\frac{q-r}{q}} T^{\ast}\right) = 
{S^{\ast}}^{\frac{q-r}{q}-1}\; \frac{1}{q}\; 
\sum_i\; \alpha_i\; 
{X^{i\; \ast}}^{-\frac{\alpha_i-r}{\alpha_i}+1}\;
d\left({X^{i\; \ast}}^{\frac{\alpha_i-r}{\alpha_i}}\; \xi_i^{\ast}\right).
\eeq
A formula analogous to (\ref{gdgw}) can be obtained if some independent 
zero-weight variables appear; one obtains (cf. Appendix \ref{gdap})
\beq
\frac{q}{r}\;  {S^{\ast}}^{-\frac{q-r}{q}+1}\; 
d\left({S^{\ast}}^{\frac{q-r}{q}} T^{\ast}\right)
+\sum_{i\leq p}\; \frac{\alpha_i}{r}\; 
{X^{i\; \ast}}^{-\frac{\alpha_i-r}{\alpha_i}+1}\;
d\left({X^{i\; \ast}}^{\frac{\alpha_i-r}{\alpha_i}}\; \xi_i^{\ast}\right)
-\sum_{p<i\leq n}\;\xi_i^{\ast}\; dX^{i\; \ast}=0. 
\eeq

\subsection{Legendre transforms}

Legendre transforming of the potentials involves the usual procedure. 
A noticeable difference, with respect to standard thermodynamics, is that, if, 
between the $n+1$ independent variables, no 
variable of weight zero appears and no intensive dependent variable 
occurs, then one can iterate the Legendre transformations  
of the potential $n+1$ times and find a potential which is 
not identically vanishing. Moreover, the would-be intensive variables 
appear in the Euler operator involving the new independent variables. 
Let us e.g. consider the quasi-homogeneous free energy 
\beq
F^{\ast}(T^{\ast},X^{1\; \ast},\ldots,X^{n\; \ast})=U^{\ast}-T^{\ast}\; S^{\ast}.
\eeq
It is a quasi-homogeneous function of degree $r$ with respect to 
the Euler operator
\beq
D_1 = (r-1)\; T^{\ast}\; \frac{\pa}{\pa T^{\ast}}+ 
\sum_i\; \alpha_i\; 
X^{i\; \ast}\; \frac{\pa }{\pa X^{i\; \ast}}.
\eeq
One can Legendre-transform also with respect to the remaining $n$ 
variables $X^{1\; \ast},\ldots,X^{n\; \ast}$:
\beq
G_{n+1}^{\ast}(T^{\ast},\xi_1^{\ast},\ldots,\xi_n^{\ast}) =
U^{\ast}-T^{\ast}\; S^{\ast}-\sum_{i=1}^{n}\; 
X^{i\; \ast}\; \xi_i^{\ast}\not \equiv 0.
\eeq
The corresponding Euler operator is 
\beq
D_{n+1} = (r-1)\; T^{\ast}\; \frac{\pa}{\pa T^{\ast}}+ 
\sum_i\; (r-\alpha_i)\; \xi_i^{\ast}\;
\frac{\pa }{\pa \xi_i^{\ast}}.
\eeq
Notice that, all the potentials which are obtained by Legendre-transforming 
the internal energy $U^{\ast}$ are quasi-homogeneous of degree $r$ 
with respect to the corresponding Euler operator; all the Massieu-Planck 
potentials, which are obtained by Legendre-transforming the entropy 
$S^{\ast}$, are quasi-homogeneous of degree one. In other terms, 
the Legendre transform preserves the degree of quasi-homogeneity 
(but it changes variables and weights: the function 
\beq
h^{\ast}(\xi^{\ast},Y^{\ast})=g^{\ast}-\xi^{\ast}\; X^{\ast}.
\eeq
which is obtained by a 
Legendre transform with respect to a variable $\xi$ of weight $\alpha$ of 
a function $g(X^{\ast},Y^{\ast})$ which is quasi-homogeneous of degree $r$ and
weights $(r-\alpha,r-\beta)$, has weight $(\alpha,r-\beta)$). 
For a proof, see Ref. \cite{hankey} 
(theorem 2, Appendix A therein).

\subsubsection{case where weight-zero independent variables appear}

Let us consider the case where a variable $\eta^{\ast}$ 
of degree $r$ and a variable $I^{\ast}$ of weight zero appears:
\beq
dU^{\ast}=T^{\ast}dS^{\ast}+\xi^{\ast}\; dX^{\ast}+\eta^{\ast}\; 
dI^{\ast}.
\eeq
The Euler operator is 
\beq
D = S^{\ast}\; \frac{\pa }{\pa S^{\ast}}+\beta\;
X^{\ast}\; \frac{\pa }{\pa X^{\ast}}.
\eeq
One can Legendre-transform three times, obtaining
\beq
G_3(T^{\ast},\xi^{\ast},\eta^{\ast})=
\frac{1}{r}\; \left[ (1-r) T^{\ast}S^{\ast}-(r-\beta)\xi^{\ast}X^{\ast} 
-r \eta^{\ast} I^{\ast}\right]
\eeq
and
\beq
D_3 = (r-1)\; T^{\ast}\; \frac{\pa}{\pa T^{\ast}}+ 
(r-\beta)\; \xi^{\ast}\;
\frac{\pa }{\pa \xi^{\ast}}+r\; \eta^{\ast}\; 
\frac{\pa }{\pa \eta^{\ast}}.
\eeq

\subsubsection{case with would-be intensive variable of degree zero}

Let us consider the case where a variable $\zeta^{\ast}$ 
of degree zero and a variable $Z^{\ast}$ of weight $r$ appears:
\beq
dU^{\ast}=T^{\ast}dS^{\ast}+\xi^{\ast}\; dX^{\ast}+\zeta^{\ast}\; 
dZ^{\ast}.
\eeq
The Euler operator is 
\beq
D = S^{\ast}\; \frac{\pa }{\pa S^{\ast}}+\beta\; 
X^{\ast}\; \frac{\pa }{\pa X^{\ast}}+r\; \zeta^{\ast}\; 
\frac{\pa}{\pa \zeta^{\ast}}.
\eeq
One can Legendre-transform three times, obtaining
\beq
G_3(T^{\ast},\xi^{\ast},\zeta^{\ast})=
\frac{1}{r}\; \left[ (1-r) T^{\ast}S^{\ast}-(r-\beta)\xi^{\ast}X^{\ast} 
\right]
\eeq
and
\beq
D_3 = (r-1)\; T^{\ast}\; \frac{\pa}{\pa T^{\ast}}+ 
(r-\beta)\; \xi^{\ast}\;
\frac{\pa }{\pa \xi^{\ast}}.
\eeq
This case is analogous to what happens in standard thermodynamics 
e.g. when one passes from $U(S,V,N)$ to $F(T,V,N)$ and then to 
$G(T,p,N)$.

\subsection{from the entropy to the energy representation and 
further considerations}
\label{enten}

In order to pass from the entropy representation $S^{\ast}(U^{\ast}, 
X^{1\; \ast},\ldots,X^{n\; \ast})$ to the energy representation 
$U^{\ast}(S^{\ast}, X^{1\; \ast},\ldots,X^{n\; \ast})$, 
one inverts the first relation with respect to 
$U^{\ast}$, which is possible because $\pa S^{\ast}/\pa U^{\ast}>0$. 
If $S^{\ast}$ has degree one and weights 
$(r,\alpha_1,\ldots,\alpha_n)$, 
then $U^{\ast}$ has degree $r$ and weights $(1,\alpha_1,\ldots,\alpha_n)$. 
Analogous 
considerations hold when passing from the energy to the entropy 
representation. We show that, when it is possible to invert, at least 
locally, a quasi-homogeneous function with respect to one variable, 
a quasi-homogeneous function is obtained again, having obvious 
degree and weights.\\
\\
In the general case, let us consider a quasi-homogeneous function 
$w=g(x^1,\ldots,x^n)$ of degree $r$ and weights $(a_1,\ldots,a_n)$; 
the partial derivatives
\beq
p_i\equiv \frac{\pa w}{\pa x^i}
\eeq
are quasi-homogeneous functions of degree $r-a_i$ for $i=1,\ldots,n$. 
Let us assume e.g. that $p_1\not = 0$. Then, at least 
locally, one can invert $w$ with respect to 
$x^1$ and find $x^1 = h(w,x^2,\ldots,x^n)$ such that 
$g(h(w,x^2,\ldots,x^n))=w$. 
The inverse function $h$ (where it exists and it is sufficiently 
smooth) is easily shown to be a quasi-homogeneous 
function of degree $a_1$ and weights $(r,a_2,\ldots,a_n)$. 
In fact, one has 
\beq
g(\lambda^{a_1} x^1,\lambda^{a_2} x^2,\ldots,\lambda^{a_n} x^n)=
g(\lambda^{a_1} h(w,x^2,\ldots,x^n),\lambda^{a_2} x^2,\ldots,
\lambda^{a_n} x^n)=
\lambda^{r}  g(x^1,\ldots,x^n).
\eeq
Moreover, one has 
\beq
g( h(\lambda^{r} w,\lambda^{a_2} x^2,\ldots,\lambda^{a_n} x^n),
\lambda^{a_2} x^2,\ldots,\lambda^{a_n} x^n)=\lambda^{r} w = 
\lambda^{r}  g(x^1,\ldots,x^n), 
\eeq
thus it has to hold 
\beq
h(\lambda^{r} w,\lambda^{a_2} x^2,\ldots,\lambda^{a_n} x^n)=
\lambda^{a_1} h(w,x^2,\ldots,x^n).
\eeq
Cf. Ref. \cite{galg}, where an analogous property is 
shown in passing from the entropy representation 
to the energy representation in standard (homogeneous) thermodynamics. 
See also \cite{landbost}.\\
\\
As a further example, 
let us consider Kerr-Newman black holes. By inverting 
(where possible) the relation $T=T(M,Q,J)$ with respect to 
$M$, one obtains (at least formally) $M=M(T,Q,J)$. 
We know that $M$ and $Q$ have degree $1/2$,  
$J$ has degree one and $T$ has degree $-1/2$. In the 
new variables, $M=M(T,Q,J)$ is a quasi-homogeneous function 
of degree $1/2$ and weights $(-1/2,1/2,1)$. 
The Euler 
operator corresponding to the independent variables $(T,Q,J)$ is 
\beq
\bar{D}=-\frac{1}{2}\; T\; \frac{\pa}{\pa T}+
\frac{1}{2}\; Q\; \frac{\pa}{\pa Q}+
J\; \frac{\pa}{\pa J}
\eeq
and it holds
\beq
\bar{D} M=\frac{1}{2}\; M.
\eeq

\section{further insights from black holes}
\label{bhins}

A feature of quasi-homogeneous thermodynamics, 
shared also with homogeneous thermodynamics, is that in the 
expression of the metrical entropy 
and of the absolute temperature a undetermined multiplicative 
constant appears (cf. also \cite{belhom}). Such a constant is 
phenomenologically fixed once the absolute scale of temperature 
is chosen. From the point of view of the actual analytic expression of 
this constant, it should be furnished by a statistical mechanical 
calculation. The black hole case can be again useful. 
By dimensional analysis, if we make the hypothesis that 
the entropy of a black hole can depend only on 
$\hbar,\; c,\; G,\; k_b$, we find that
\beq
S\propto\; \frac{k_b\; c^3}{\hbar\; G}\; A; 
\eeq
this is the same procedure as 
in the original proposal of Bekenstein \cite{beken}. 
For the temperature one obtains
\beq
T\propto\; \frac{\hbar\; c}{k_b}\; 
\left(\frac{\pa A}{\pa\; (G M/c^2)}\right)^{-1}.
\label{hawkin}
\eeq
A dimensionless constant has still to be determined.\\ 
If one considers standard thermodynamics cases, 
this hypothesis which allows to recover the dimensions of the entropy by 
involving only the above constants works for the photon gas but 
it does not work in other cases; e.g. in the case of a Fermi gas of 
electrons, the electron mass represents a further scale, related to 
a microscopic analysis of the system, to be taken into account. 
In our case, the confirmation of the 
above hypothesis comes from the Hawking effect, which gives 
us the behavior of the black hole temperature, to be compared 
with (\ref{hawkin}). Then, because of the absence of a particle mass 
scale and, e.g. in the Schwarzschild and in the Kerr cases being involved 
only vacuum solutions of the 
Einstein equations, one could be tempted to think about some 
sort of ``graviton gas'', but, actually, 
this is far from being evident. See also Ref. \cite{belcar}. 

Black holes represent 
an example of the the procedure of reducing/enlarging consistently 
the thermodynamics space discussed in sect. \ref{general}. 
In fact, the construction of the fundamental  
equation sketched in sect. \ref{bhtd} 
is allowed also in the case of a black hole 
with $J=0$ and of a black hole with $Q=0$. Two parameters appear, 
$M,Q$ and $M,J$ respectively. The integrability condition 
is then trivially satisfied, but a well-defined potential $S$ 
is obtained everywhere. Moreover, 
black holes of the Kerr-Newman family are a consistent extension 
of the black holes of the Reissner-Nordstr\"{o}m family 
(by adding $J$) or of the Kerr family (by adding $Q$). 
To this family belong also a thermodynamically 
degenerate case: the Schwarzschild case $J=0=Q$, which is 
described only by one variable $M$. Notice also that the Kerr-Newman family 
can be obtained by putting $\Lambda=0$ in the KN-AdS family.

In the case of 
black hole thermodynamics, it is tempting to conjecture that 
the quasi-homogeneous behavior of the thermodynamic potentials 
could be related to  
an explanation, to some extent, analogous to the one for the 
quasi-homogeneous behavior of the thermodynamic potentials 
in standard thermodynamics near the critical point \cite{uzunov,henkel}: 
As in standard 
thermodynamics the (leading order) quasi-homogeneous behavior 
can be related to the conformal invariance of the underlying 
quantum field theory near criticality \cite{henkel}, one 
could conjecture that the quasi-homogeneous behavior 
of black hole thermodynamics could be related to some sort 
of conformal invariance of the quantum  theory 
underlying General Relativity (superstring theory?). 
This field deserves further investigation. 
We however point out that it does not seem to be 
necessary that  
quasi-homogeneous thermodynamics should be related 
to some sort of conformal invariance in every case. For 
black holes, as 
well as for the general case, one should determine the 
reason of a quasi-homogeneous behavior by means of statistical 
mechanical calculations.  In the following section, 
a further discussion is found.

\section{Heuristics: Quasi-homogeneous thermodynamics and thermodynamic limit}
\label{heutd}

In this section we suggest that the thermodynamic limit, under 
suitable hypotheses, can lead only to quasi-homogeneous thermodynamics. 
The thermodynamic limit should not be 
intended in a literal sense, of course, but as a tool which allows to 
determine the leading terms in the statistical mechanical  
functionals as large systems occur. Cf. e.g. Ref. \cite{munster}, chap. 4, 
and also Ref. \cite{still}. Thermodynamics of macro-systems emerges 
as an asymptotic law which is extrapolated from the asymptotic 
behavior of statistical mechanics. To some extent, the thermodynamic 
limit is also a tool by means of which the vague notion of ``macroscopic 
system'' is meant to be implemented, and a proper thermodynamic behavior 
recovered. 
Such a behavior depends 
on the interactions and on the 
nature of the system involved (e.g., fermionic or bosonic matter), 
together with a dependence on the initial conditions [from a 
quantum-mechanical point of view, a sort of ``preparation'' of the 
system, even if it is of ``astrophysical size'']. 
As far as the gravitational interaction can be neglected, the standard 
thermodynamic behavior emerges, with the usual asymptotic laws which  
justify the extensivity of standard thermodynamics. At the same time, 
extensivity cannot be considered as an absolute and unmodifiable property of 
thermodynamics, because of the purely attractive nature of gravity. 
One can also 
expect a different asymptotic behavior for different systems at 
different scales. See also the discussion in the following subsections.   
\\
The basic set (B) of assumptions is the following:\\
\\
\underline{Assumptions (B)}:\\
{\sl  
Let $\{ Z_i \}$ be the set of the  
thermodynamic variables, both independent ($Z^{(I)}_i$) 
and dependent ($Z^{(D)}_i$) ones. Let the independent variables 
describe the Gibbs thermodynamic space.\\
Let $\lambda$ be the parameter such that, by 
taking $\lambda\to \infty$, one recovers the thermodynamic limit.\\   
Assume that the thermodynamic limit properly describes the 
bulk properties of the system under consideration. \\ 
Assume that the thermodynamic limit of the statistical mechanical 
quantities, like e.g. the entropy or the free energy, 
exists under the hypothesis that, in the 
limit $\lambda\to \infty$, the independent 
thermodynamic variables $Z^{(I)}_i (\lambda)$ are rescaled as
\beq
Z^{(I)}_i= g^{(I)}_i (\lambda)\; Z^{(I)\ 0}_i,
\eeq
where $g^{(I)}_i (\lambda)$ is a positive function which is 
either an invertible function  
of $\lambda$ or it is one (or a constant),
and $Z^{(I)\ 0}_i$ is the 
value of the $i$-esime variable $Z^{(I)}_i$ at an arbitrary reference state 
(cf. Ref. \cite{still} for the extensive case).}\\
\\
We now state the following result (R-group).\\
\\
%\newpage 
\noindent
\underline{Result (R-group)}:\\
{\sl 
Let us assume the set of assumptions (B) and 
that for the rescalings of the independent 
variables the group property holds:
\beq
g^{(I)}_i (\lambda \mu)=g^{(I)}_i (\lambda)\; g^{(I)}_i (\mu).
\label{gprop}
\eeq 
This implies that for each independent variable there 
exists a real number $\alpha^{(I)}_i$ such that 
$g^{(I)}_i (\lambda)=\lambda^{\alpha^{(I)}_i}$.\\ 
Let us assume that the dependent thermodynamic variables 
\beq
Z^{(D)}_i=Z^{(D)}_i (Z^{(I)}_1,\ldots,Z^{(I)}_n),
\eeq
are obtained as leading terms in the asymptotic expansion of 
the corresponding statistical mechanical 
functionals ${\cal Z}^{(D)}_i$
in the following sense: for each $i$ there exists a positive continuous 
function $\rho_i (\lambda)$ such that 
\beq
\lim_{\lambda\to \infty}\; \frac{1}{\rho_i (\lambda)}\; {\cal Z}^{(D)}_i 
(g^{(I)}_1 (\lambda)\; Z^{(I)}_1,\ldots, g^{(I)}_n (\lambda)\; Z^{(I)}_n)=
Z^{(D)}_i (Z^{(I)}_1,\ldots,Z^{(I)}_n),
\eeq
and that the domain of the functionals ${\cal Z}^{(D)}_i$ is invariant 
under the above rescalings.\\ 
Then, the dependent thermodynamic variables are quasi-homogeneous, 
i.e. they satisfy
\beq
Z^{(D)}_i (g^{(I)}_1 (\lambda)\; Z^{(I)}_1,\ldots, g^{(I)}_n (\lambda)\; Z^{(I)}_n)=g^{(D)}_i (\lambda)\; Z^{(D)}_i (Z^{(I)}_1,\ldots,Z^{(I)}_n)
\eeq
where $g^{(D)}_i (\lambda)=\lambda^{\alpha^{(D)}_i}$ and 
$\alpha^{(D)}_i\in \RR$.\\ 
Moreover, 
for all $i$, the functions $\rho_i (\lambda)$ are regularly varying functions 
\cite{seneta}:\\
for all $\mu\in \RR_+$\\
\beq
\lim_{\lambda\to \infty}\; \frac{\rho_i (\lambda\; \mu)}{\rho_i (\lambda)}=
\mu^{\alpha^{(D)}_i}. 
\eeq  
($\alpha^{(D)}_i$ is also called degree of $\rho_i$) }.\\
\\ 
In order to be more explicit, let us consider (R-group) for 
a specific case. 
Let us e.g. assume that 
the thermodynamic limit for the statistical mechanical functional 
${\cal S}$ representing the entropy exists in the sense that 
\beq
\lim_{\lambda\to \infty}\; \frac{1}{\rho_S (\lambda)}\; {\cal S}
(\lambda^{\alpha_U}\; U, \lambda^{\alpha_V}\; V, \lambda^{\alpha_N}\; N)=
S(U,V,N),
\label{lien}
\eeq
where the function $\rho_S$ is positive and 
$\alpha_U,\alpha_V,\alpha_N$ are real numbers.  
Then for each real $\mu>0$ the function $\rho_S$ satisfies 
\beq
\lim_{\lambda\to \infty}\; \frac{\rho_S (\lambda\; \mu)}{\rho_S (\lambda)}=
\mu^{\gamma}
\eeq
for some $\gamma \in \RR$. Moreover, $S$ is quasi-homogeneous of 
degree $\gamma$. In fact, from (\ref{lien}) it follows 
\beq
\lim_{\lambda\to \infty}\; \frac{1}{\rho_S (\lambda\; \mu)}\; {\cal S}
((\lambda\; \mu)^{\alpha_U}\; U, (\lambda\; \mu)^{\alpha_V}\; V, 
(\lambda\; \mu)^{\alpha_N}\; N)=S(U,V,N).
\label{lienmu}
\eeq
Moreover, it holds 
\beq
\lim_{\lambda\to \infty}\; \frac{1}{\rho_S (\lambda)}\; {\cal S}
(\lambda^{\alpha_U}\; (\mu^{\alpha_U}\; U), 
\lambda^{\alpha_V}\; (\mu^{\alpha_V}\; V), 
\lambda^{\alpha_N}\; (\mu^{\alpha_N}\; N))=
S(\mu^{\alpha_U}\; U,\mu^{\alpha_V}\; V, \mu^{\alpha_N}\; N).
\eeq
Then 
\beqnl
&&\lim_{\lambda\to \infty}\; \frac{\rho_S (\lambda)}{\rho_S (\lambda\; \mu)}\; 
\frac{{\cal S}
((\lambda\; \mu)^{\alpha_U}\; U, (\lambda\; \mu)^{\alpha_V}\; V, 
(\lambda\; \mu)^{\alpha_N}\; N)}{{\cal S}
(\lambda^{\alpha_U}\; (\mu^{\alpha_U}\; U), 
\lambda^{\alpha_V}\; (\mu^{\alpha_V}\; V), 
\lambda^{\alpha_N}\; (\mu^{\alpha_N}\; N))}=\cr
&&\frac{S(U,V,N)}{S(\mu^{\alpha_U}\; U,\mu^{\alpha_V}\; V, \mu^{\alpha_N}\; N)}
=\lim_{\lambda\to \infty}\; \frac{\rho_S (\lambda)}{\rho_S (\lambda\; \mu)},
\eeqnl
which is possible only if the above conditions are implemented. 
A general proof is found in Appendix \ref{asyap}. Cf. \cite{vladimirov}, 
where a rigorous approach for asymptotics of tempered distributions 
is developed.\\  
The case where one or more variables are ``intensive'', i.e. are not 
rescaled, presents no difficulty. The above reasoning about the 
leading order in the asymptotic law identified in the thermodynamic 
limit is corroborated by the standard thermodynamic limit, which 
allows to find a leading-order homogeneous thermodynamics. We 
define {\sl quasi-homogeneous thermodynamic limit} the thermodynamic limit 
satisfying the group property.\\
Moreover, 
we notice that, if e.g. for the thermodynamic 
entropy $S(U,V,N)$ it holds 
\beq
S(g_U (\lambda)\; U_0, g_V (\lambda)\; V_0, g_N (\lambda)\; N_0)=
g_S (\lambda)\; S_0=
g_S (\lambda)\; S(U_0,V_0,N_0),
\label{sho}
\eeq
where the functions $g_U, g_V, g_N, g_S$ are positive and  
the group property holds, then 
$S$ has to be quasi-homogeneous and the $g_i$ are power-like functions. 
The key point is that 
a behavior like the one 
in (\ref{sho}) under the group property  is necessarily quasi-homogeneous;  
as it is shown in Ref. \cite{hankey}, ``powers need no generalization'' 
[see Appendix A in Ref. \cite{hankey} and Appendix C in Ref. 
\cite{chang}]. A different proof is sketched in Appendix  \ref{qoap} for the 
sake of completeness. 
The aforementioned theorem implies that actually one has
\beq
S(g(\lambda)^{\alpha_U}\; U, g(\lambda)^{\alpha_V}\; V,
g(\lambda)^{\alpha_N} N)=
g(\lambda)^{\alpha_S}\; S(U,V,N)
\eeq
with the same scaling function $g(\lambda)$. By defining 
\beq
\bar{\lambda}\equiv g(\lambda)
\eeq
one finds $S(\bar{\lambda}^{\alpha_U}\; U, \bar{\lambda}^{\alpha_V}\; V,
\bar{\lambda}^{\alpha_N} N)=
\bar{\lambda}^{\alpha_S}\; S(U,V,N)$, i.e. the standard definition 
of quasi-homogeneous behavior for a function. 
A generalization of the 
quasi-homogeneous behavior of a function under rescaling with 
the group property is not allowed. 

We now relax the requirement for the group property, one can still 
make a conjecture (C-reg) which would allows a 
quasi-homogeneous behavior 
under rather general conditions.\\
\\
%\newpage
%\noindent
\underline{Conjecture (C-reg)}:\\
{\sl Let us assume the set (B) and that 
$g,g_i$ (for $i=1,\ldots,n$) are positive and continuous functions 
which are regularly varying 
functions of real degree $\gamma, \alpha_1,\ldots,\alpha_n$ respectively.\\ 
Assume that 
the statistical mechanical functional ${\cal Z}$ 
satisfies
\beq
\lim_{\lambda\to \infty}\; \frac{1}{g(\lambda)}\; 
{\cal Z}( g_1 (\lambda)\; Z^1,\ldots, g_n (\lambda)\; Z^n)=
Z(Z^1,\ldots,Z^n),
\label{limhyp}
\eeq
where $Z^i$ are the independent 
thermodynamic variables (we omit the suffix (I)).\\ 
We conjecture that, under suitable hypotheses, 
the asymptotic $Z(Z^1,\ldots,Z^n)$ 
is quasi-homogeneous of degree $\gamma$ and weights 
$(\alpha_1,\ldots,\alpha_n)$.}\\
\\
For each positive real $\mu$ one should find
\beqnl
&&\lim_{\lambda\to \infty}\; \frac{1}{g(\lambda\; \mu)}\; 
{\cal Z}( g_1 (\lambda\; \mu)\; Z^1,\ldots, g_n (\lambda\; \mu)\; Z^n)=
Z(Z^1,\ldots,Z^n)\\
&&``=\hbox{''}\lim_{\lambda\to \infty}\; \frac{1}{g(\lambda)\; \mu^{\gamma}}\; 
{\cal Z}( g_1 (\lambda)\; \mu^{\alpha_1}\; Z^1,\ldots, 
g_n (\lambda)\; \mu^{\alpha_n}\; Z^n)\\
&&=\frac{1}{\mu^{\gamma}}\; 
Z(\mu^{\alpha_1}\; Z^1,\ldots,\mu^{\alpha_n}\; Z^n)
\eeqnl
where ``$=$'' indicates that further hypotheses on the 
statistical mechanical functional should be satisfied. Notice that the case 
$g_i (\lambda)=\lambda^{\alpha_i}$ is a sub-case of the preceding one and 
requires only that the limit (\ref{limhyp}) exists and that the state 
$(\mu^{\alpha_1}\; Z^1,\ldots, \mu^{\alpha_n}\; Z^n)$ 
belongs to the domain of the statistical mechanical functional ${\cal Z}$.\\
\\
\\
From a physical point of view, it can be noticed that, when passing from 
the statistical mechanical functional ${\cal S}$ representing the entropy 
to its asymptotic $S_{sm}$ under dilatations, as in the standard 
thermodynamic limit, it would be not strictly 
necessary, a priori, to impose that $g (\lambda)=\lambda$ in order 
to obtain an homogeneous  $S_{sm}$ 
[herein we indicate by $S_{sm}$ the 
asymptotic which is indicated by $S$ in (\ref{lien}). $S_{sm}$ corresponds 
to the statistical mechanical entropy one takes into account in 
sect. \ref{stawei}]. 
A generic regularly varying function of 
degree one could also be allowed (e.g., $g(\lambda) = 
\lambda\; \log(\lambda)$ could be allowed). The requirement 
\beq
\lim_{\lambda \to \infty}\; \frac{1}{\lambda}\; {\cal S}(\lambda \; x)=
S_{sm}(x)
\eeq
amounts to the requirement that in some sense also ${\cal S}$ is extensive.\\ 
We have found that, under rather general hypotheses, the only possible 
outcome of the thermodynamic limit is a form of quasi-homogeneous 
thermodynamics. 
The thermodynamic limit, at the same time, 
is not expected to describe all possible systems; e.g., it cannot be 
applied to a thin (molecular thickness) metallic film \cite{balescu}. 
Nevertheless, it 
can be safely applied to a huge class of macroscopic systems, and, in the 
non-relativistic case, it allows to conclude that 
the Thomas-Fermi approximation becomes exact in the limit.\\
In partially concluding these heuristic considerations, we 
point out that a relation between 
a so-called pseudo-extensive thermodynamics and the thermodynamic 
limit  has been postulated in Ref. \cite{veguz} and an analysis of 
Newtonian self-gravitating systems is made in Ref. \cite{vegrav}. 
We limit ourselves to refer the reader to the aforementioned papers. 

\subsection{non-conventional thermodynamic limit}

A non-conventional thermodynamic limit has already been proposed. 
A very interesting discussion of the thermodynamic limit in physical 
systems is found in Ref. \cite{kiessling}. Therein, physical systems 
are divided into two classes each of which can be associated with its own 
thermodynamic limit; one can distinguish between the standard thermodynamic 
limit (STL) and the inhomogeneous mean field thermodynamic limit 
(IMFTL) which occurs e.g. in self-gravitating systems as the one 
of HNT model. The latter kind of thermodynamic limit can be related with 
the presence of long-range forces \cite{kiessling}. In order to identify 
the correct limit, the key notion of characteristic bulk length scale 
is introduced, which corresponds to a typical length scale $\lambda_{typ}$ 
resulting from the interaction of many particles in the system. 
Such length scale $\lambda_{typ}$ plays the role of characteristic 
invariant in the thermodynamic limit. In the case of a 
classical self-gravitating isothermal gas, the Jeans length $\lambda_{Jeans}$ 
is such a typical length. 
Moreover, the requirement that 
mean thermodynamic quantities exist almost everywhere in the thermodynamic 
limit is a tool for selecting the appropriate thermodynamic limit 
at constant $\lambda_{typ}$ \cite{kiessling}.  
A complementarity between STL and IMFTL can also be allowed for 
the same system, in the sense that STL and IMFTL can describe 
complementary asymptotic properties for the same large but finite system 
(the case of classical Coulomb systems is discussed). We don't discuss 
further on this approach herein, we limit ourselves to point out that 
one obtains the same limit, in the case of classical self-gravitating 
matter, if one keeps $\lambda_{Jeans}$ fixed or $\xi$  
fixed (cf. sect. \ref{hnt}). Notice that $\xi$ is an intensive function 
with respect to the Euler operator relative to the HNT model. 
At an heuristic level one expects that in general there exists 
a physically meaningful function which is left 
invariant under a quasi-homogeneous thermodynamic limit and 
which is intensive under the corresponding Euler operator. 

A non-conventional 
discussion of the thermodynamic limit is found also in 
sections 1.2 and 4.2 of Ref. \cite{thirbook}.  
Both for stable and for 
unstable interactions are discussed, and the criterion for selecting the 
limit is a comparison between the kinetic and the potential energies 
involved in the system. According to \cite{messer} 
[p. 1-3], a  non-conventional thermodynamic limit has to be associated 
with  the peculiar scaling behavior for ``non-stable interactions'';  
the notion of ``Thomas-Fermi thermodynamic limit'' 
is also introduced for non-extensive systems.  
Note that a more general thermodynamic limit 
can also be allowed, where also the coupling constants are allowed to 
vary in the limit. Cf. \cite{hertel,messer,kiessling}.

\subsection{absolute weights again}

It is evident that one has a reference quantity 
with respect to which all the others are ``weighted'' in the thermodynamic 
limit. Then an overall undetermined scale factor $q$ is expected to 
appear, unless an absolute weight is known.\\ 
In the standard homogeneous case, the problem of assigning absolute weights 
does not arise, they are all equal to one for extensive 
variables and zero for intensive ones, the thermodynamic limit 
is performed with the aim of studying bulk properties of the system, 
and the ratios $U/N$ and $V/N$ are kept constant. Roughly, one fixes 
the energy per particle and the volume per particle (or, equivalently, 
the density $N/V$ and the energy density $U/V$) and then performs 
the limit with the aim of neglecting any finite-size effect. 
The same is true also for a massless particle gas, like 
the photon gas, where the limit $U,V\to \infty, U/V =$ const. is 
performed.\\ 
If a strict quasi-homogeneous picture is required, then the above 
problem is instead present. When $N$ is available, 
as it is when non-zero rest-mass particles 
are considered, it is to some extent natural to assume that its weight 
is one.  
Such a special role of $N$ can be justified. 
In the model of Hertel, Narnhofer and Thirring, 
the existence of the Boltzmann's entropy per particle in the 
thermodynamic limit $\lim_{N\to \infty} \log(\Omega)/N$ is ensured, 
in agreement with 
Boltzmann's postulate, which states that the entropy per particle 
$\log(\Omega)/N$ exists for an equilibrium system. As a consequence of 
the coincidence of  $\lim_{N\to \infty} \log(\Omega)/N$ with the 
thermodynamic entropy density (per particle), one can justify a posteriori 
the assumption that $N$ has weight one, in fact $\log(\Omega)$ in the 
thermodynamic limit coincides with the thermodynamic entropy which 
has degree one. 
The other weights $\alpha_i$ appropriate to the physical condition one 
is considering for the system are recovered in the asymptotic limit.
Then, no ambiguity for an overall multiplicative constant in the 
definition of the weight like the one discussed in sect. \ref{ambig} 
appears. 
Note also that the Boltzmann's entropy, being the logarithm of the 
number of micro-states compatible with a given macro-state, by involving 
the (vague) concept of macro-state and thus of macroscopic system 
can legitimately related to the thermodynamic limit (again in the sense 
of asymptotic law). [About Boltzmann's entropy and its coincidence 
with the Gibbs entropy (which, in the 
quantum case, corresponds to the 
von Neumann entropy) see e.g. \cite{goldstein}.] 
A more puzzling case is represented by a 
self-gravitating massless gas, but weights can be still 
fixed. See sect. \ref{geon}.
 
In general, if $\lambda\to \infty$ corresponds to the thermodynamic 
limit, one could also require the existence of the Boltzmann entropy 
density functional as $\lambda \to \infty$, to be defined as 
\beq
s_{boltz}\equiv 
\lim_{\lambda \to \infty} \frac{1}{\lambda^{\alpha_S}} \log(\Omega),
\label{ebolt}
\eeq
where $\alpha_S$ is a constant which is fixed by the requirement 
of finiteness of the above limit. $\alpha_S$ has to be introduced 
because the scale $\lambda$ can also be such that 
$\lambda/N$ is not asymptotically a positive constant, i.e., 
$\lambda \not \sim N$ in the sense of the asymptotic behavior. 
This more general choice for the scale 
is also necessary if $N$ is not a good thermodynamic variable 
(as in the massless gas case). We know that it is consistent to 
fix the weights by choosing the Boltzmann entropy to be of degree one.\\
If another variable, e.g. the internal energy, 
would be used as reference variable, the relative weights could be 
still recovered in line of principle, but the aforementioned 
ambiguity would emerge in absence of a criterion for finding the 
absolute weight at least of one (independent or dependent) variable.\\
 
One could also choose to work in the energy representation, and 
decide to fix the weight of the internal energy to be one. Then the 
entropy in general would not be a degree one quasi-homogeneous function; 
in the HNT model, one would obtain 
deg$(V)=-3/7$; deg$(N)=3/7$; deg$(S)=3/7$, against the weight $7/3$ of 
$U$ resulting by assuming that $N$ has weight one. 
This procedure, although legitimate because of the overall ambiguity, 
is less satisfactory, because the thermodynamic construction, in the case 
the entropy is a quasi-homogeneous function and the weights present 
the overall factor ambiguity, allows naturally the thermodynamic 
entropy to be of degree one. The only exception which, in line of principle 
could be admitted, is given by the case where the weights in the 
Gibbs space are all fixed and the statistical-mechanical entropy 
is not of degree one, but it is apparently not allowed (unless 
some reason for fixing the absolute weight of a variable appears).

\subsection{thermodynamics, weights and gravity}

As far 
as gravity is negligible, one finds the standard extensivity property 
for the fundamental equation in the entropy [energy] representation; 
when the scale of the system is such that is no more possible 
to neglect self-gravity effects, 
then the weights of the thermodynamic 
variables change according to the nature of the system and to 
the scale itself. The quasi-homogeneous character of thermodynamics 
with fixed weights $(\alpha_1,\ldots,\alpha_n)$ 
has to be intended as relative to the validity 
range of the asymptotic expansion which leads to those 
particular $(\alpha_1,\ldots,\alpha_n)$, whose nature is 
thus non absolute. [One could also be tempted to write $\alpha_i = 
\alpha_i (\lambda)$, in order to recall that the weights can change 
with the scale. Nevertheless, in light of the above theorem about 
quasi-homogeneity, such a dependence is not allowed. The quasi-homogeneity 
is extrapolated at each fixed scale, and it is made ``absolute'' by 
the thermodynamic formalism.]
Also an estimate of the error for finite $\lambda$ should 
be allowed \cite{munster}. One could think also to a sort of 
``evolving picture'' for the formalism, in which one starts from 
standard homogeneous systems, characterized by stable interactions, 
and then, as far as gravity becomes non-negligible, a quasi-homogeneous 
behavior is allowed. 
Weights can change, even in 
a discontinuous way, depending on the physical conditions leading 
to the new stationary equilibrium state. 
Because of the Hawking effect, the 
``end of the story'' does not occur even when the scale and the physical 
conditions are  such 
that a black hole forms, whose ``classical'' weights are the ones 
presented in sect. \ref{bhtd}.\\ 
It is interesting to underline again that ordinary thermodynamics 
implements the picture described above, as discussed in 
several fundamental papers on the stability of matter 
\cite{fisher,dyson,leblond,lieblebo}. See also Refs. \cite{ruelle,liebook}. 
Strict quasi-homogeneity is expected to 
be effective when gravitation begins playing an essential role 
in the physics of the system. Moreover, the 
so-called fourth law of thermodynamics \cite{landstat} should be 
generalized in such a way to include the general quasi-homogeneous 
behavior of the thermodynamic variables [dependent as well as independent 
ones].

\section{conclusions}

Quasi-homogeneous thermodynamics is proposed as 
the unifying picture for thermodynamics where both stationary 
black holes, fermionic 
non-relativistic matter, 
and self-gravitating electromagnetic radiation are found.\\
We have chosen to introduce quasi-homogeneity in thermodynamics 
by generalizing the formalism developed for standard 
homogeneous thermodynamics \cite{belhom}.   
The basic requirement is that the Euler vector field which 
generates quasi-homogeneous dilatations in the Gibbs' thermodynamic 
space is a non-trivial symmetry for the integrable Pfaffian form $\deq$. 
This means that the symmetry maps leaves of the thermodynamic foliation 
onto other leaves. The leaves of the foliation are naturally associated with 
a quasi-homogeneous function $S^{\ast}$ of degree one, which is 
suggested to be the (unique apart from a multiplicative constant) 
metrical entropy corresponding to the 
fundamental equation in the entropy representation. This is true 
also in the case one knows the weights of the thermodynamics variables 
apart from an overall unknown factor. Notice that quasi-homogeneity 
could be introduced also at the level of the fundamental equation 
in the entropy representation or in the energy one, but, in the 
case where only the relative weights of the independent 
thermodynamic variables are known, it should be considered 
equivalent the choice of a degree one internal energy with respect 
to the choice of a degree one entropy (cf. also \cite{hankey}), whereas 
the aforementioned approach privileges a degree one entropy. 

The consequences of the lack 
of homogeneity are recalled, and 
superadditivity of the metrical entropy is a privileged property 
ensuring the second law even in absence of concavity for $S^{\ast}$.
 
The generalized Gibbs-Duhem identities are also derived, and 
the energy representation and the behavior under Legendre transforms 
is analyzed.

An heuristic argument which can relate quasi-homogeneous thermodynamics 
to the thermodynamic limit in statistical mechanics (even if in a 
non-conventional framework) has been introduced. 
According to this argument, under the group property 
only a form of quasi-homogeneous 
thermodynamics can be recovered in the thermodynamic limit. Moreover, 
a mean field thermodynamic limit appears to be appropriate; the 
suggestion from the existing models is that one has to search for 
asymptotic scaling properties of the equations allowing to find out the 
mean field thermodynamics for the system, in fact the 
scaling properties of a finite 
temperature Thomas-Fermi equation appears to be involved in the 
quasi-homogeneity of the HNT model, whereas the scaling 
properties of the TOV equations are related with the quasi-homogeneity 
of self-gravitating radiation. 
\\  
Gravity appears to play a fundamental role in allowing a 
generalization of the standard scheme for thermodynamics. 
The purely attractive nature of gravity,  
as well-known, is the reason for the failure of the 
extensivity property in thermodynamics of self-gravitating 
systems, because of the absence of saturation which leads to 
an implosive instability, and  matter systems eventually can 
implode into a black hole if the mass is sufficiently large. 
Moreover, 
the thermodynamic ensembles become inequivalent \cite{messer}, and 
the lack of concavity leads to phase transitions in matter systems 
in the canonical ensemble \cite{thirbook,messer}. Negative heat 
capacities arise in the micro-canonical ensemble.  
Also black holes present this feature.\\ 

Black hole thermodynamics requires some special comments. 
From the point of view of thermodynamic formalism, 
black hole thermodynamics can be realized to be a rather straightforward 
generalization of the formalism developed for standard thermodynamics 
in \cite{belhom}; moreover, its quasi-homogeneous behavior can no more 
be considered as exceptional, because self-gravitating systems 
exist where quasi-homogeneous scaling laws are satisfied.  
Nevertheless, from other points of view black hole thermodynamics is still 
to be considered special, also because it is still unclear which 
statistical mechanics should lie beyond it.\\ 

We limit ourselves herein to this proposal for a generalization of 
the thermodynamic formalism,  
without pursuing it further on [apart from a note on the third 
law in Appendix \ref{thirdap}]. 
For future investigations,    
it would be interesting to develop a statistical mechanical formalism 
where General Relativity is self-consistently included. 
\\
It would be 
also interesting to explore,  
if other long range forces are allowed to be included in 
the framework of quasi-homogeneous thermodynamics. Classical 
two-component Coulomb matter seems to belong to this framework 
\cite{kiessling}, further investigations are required in order to 
see if this is true also for quantum Coulomb matter. 

\newpage
\appendix

\section{ 
integral of $\omega/\mu$ and potentials associated with 
quasi-homogeneous exact Pfaffian forms}
\label{potap}

We first present some results in the general case of a 
generic symmetry $X$ for a Pfaffian form $\omega$. Then we 
translate the results for the case of the quasi-homogeneity symmetry. 
We recall that a vector field $X$ is defined to be a symmetry for a 
Pfaffian form $\omega$ if 
\beq
(L_X \omega)\wedge \omega = 0.
\eeq
This means that there exists a function $h$ such that 
\beq
L_X \omega = h\; \omega.
\eeq
See \cite{bocharov,cerveau}.\\ 
\\
We need the following two lemmas:\\
\\
%\newpage
%\noindent
\underline{lemma 1}\\
\\
{\sl Let $\bar{\omega}$ 
be an exact  
Pfaffian form;\\
let $X$ be a symmetry such that 
\beq
L_X \bar{\omega} = q\; \bar{\omega},
\eeq
where $q\not = 0$ is a constant. 
Then, one finds that $\bar{\omega}=d\phi$ is implemented 
by}  
\beq
\phi \equiv \frac{1}{q}\; i_X \bar{\omega}.
\label{esax}
\eeq
\underline{proof}: The proof is elementary. One has trivially
\beq
d\phi = \frac{1}{q}\; d(i_X\; \bar{\omega}) = 
-\frac{1}{q}\; i_X (d\bar{\omega})+\frac{1}{q}\; L_X\; \bar{\omega}.
\eeq
From $d\bar{\omega}=0$ because $\bar{\omega}$ is closed and from the 
symmetry of $\bar{\omega}$, which implies 
$L_X\; \bar{\omega} = q\; \bar{\omega}$ the thesis follows. Notice that 
$\phi$ satisfies $X\phi = q\; \phi$.\quad  $\square$\\
\\
\\
%\newpage
%\noindent
\underline{lemma 2}\\
\\
{\sl Let $\omega_{(0)}$ be an exact  
Pfaffian form;\\
let $X$ be a symmetry such that 
$L_X \omega_{(0)} = 0$;\\ 
let $\hat{W}$ be the 
associated potential: $d\hat{W}\equiv \omega_{(0)}$.\\
Then 
\beq
d\hat{W}=\frac{dF}{F}
\eeq
where $F$ satisfies 
\beq
X\; F = q\; F, 
\label{eXq}
\eeq
with $q\in \RR$ constant. Moreover, 
\beq
i_X\; \omega_{(0)}=q.
\eeq
As a consequence, 
$q=0$ is allowed if and only if the symmetry is tangent (i.e. trivial).}\\
\\
\underline{proof}: We have 
\beq
0=L_X \omega_{(0)}=L_X\; d\hat{W}=d\; L_X\; \hat{W}.
\eeq
As a consequence, 
\beq
L_X\; \hat{W} = X\; \hat{W} = q
\label{eqY}
\eeq
where $q$ is a constant. We can define a 
positive definite function $F$ such that 
\beq
\hat{W}\equiv \log(F).
\eeq
Then we get that (\ref{eqY}) is equivalent to the 
following equation for $F$:
\beq
X\; F=q\; F;
\eeq   
moreover
\beq
d\hat{W}=\frac{dF}{F},
\eeq
and 
\beq
i_X\;  \omega_{(0)}=i_X\; d\hat{W}=L_X\; \hat{W}=\frac{1}{F}\; X\; F=q.
\eeq
If the symmetry is tangent, then $i_X\;  \omega_{(0)}\equiv 0$. 
Then $q=0$ necessarily. If $q=0$, then $F$ and $\hat{W}$ 
satisfy $X\; F =0 $ and $X\; \hat{W}=0$ respectively. 
As a consequence, 
$i_X\;  \omega_{(0)}=i_X\; d\hat{W}=L_X\; \hat{W}=0$, 
which completes the proof.\quad $\square$\\
Notice that $\hat{W}$ has to be found by quadratures, the 
contraction of the Pfaffian form with the vector field $X$ is 
not useful in order to find a potential without explicit integration.\\  
%\\
%\newpage

One can in general show that the following theorem holds:\\
\newpage
%\\
\noindent
\underline{theorem 1}\\
\\
{\sl Let $\omega$ 
be a integrable Pfaffian form which is defined in a connected, 
simply connected domain;\\
let $X$ be a non-trivial symmetry for $\omega$;\\
let $\mu=i_X \omega$.\\ 
Then $\mu$ is an integrating factor for $\omega$ and 
the foliation corresponding to the Pfaffian equation 
$\omega=0$ is associated with a potential $F$ which satisfies 
the following equation:
\beq
X\; F = F.
\label{exf}
\eeq
}
\underline{proof}:  
The proof that $\omega/\mu$ is closed is trivial. 
See also \cite{cerveau} and \cite{belhom}. 
Moreover 
\beq
L_X\; \frac{\omega}{\mu}=0,
\eeq
being $X$ a symmetry for $\omega$. In fact, 
\beqnl
L_X\; \frac{\omega}{\mu}&=&\frac{1}{\mu^2}\left( (L_X \omega) \mu - 
(L_X \mu) \omega \right)= \frac{1}{(i_X \omega)^2}\left( (L_X \omega) 
(i_X \omega) - 
(L_X i_X \omega) \omega \right)\\
&=&\frac{1}{(i_X \omega)^2}\; i_X (\omega \wedge (L_X \omega)),
\eeqnl
where the rules $i_X L_X=L_X i_X$ and $i_X (\alpha \wedge \beta) = 
(i_X \alpha)\wedge \beta+ (-)^{n}\; \alpha \wedge (i_X \beta)$ for 
a $n$-form $\alpha$ and a $m$-form $\beta$ are used. Cf. \cite{morita}.\\
As a consequence, $\omega/\mu$ is an exact one-form 
such that the previous lemma applies.  One has
\beq
\frac{\omega}{\mu}=d\hat{W}=\frac{dF}{F},  
\eeq
where $F$ satisfied eqn. (\ref{eXq}) with $q\not =0$, 
because of the hypothesis of non-trivial symmetry. 
Then one finds 
\beq
dF = \frac{F}{\mu}\; \omega,
\label{fome}
\eeq
which satisfies $L_X\; dF = q\; dF$ and is closed. 
Then, from (\ref{esax}), we obtain 
\beq
F = \frac{1}{q}\; \frac{F}{\mu}\; i_X \omega,
\eeq
which, because of the definition of $\mu$, implies $q=1$.\quad  $\square$ 
\\
Notice that the integrating factor satisfies
\beq
L_X\; \mu = L_X\; i_X \omega = i_X\; L_X\; \omega = h\; \mu.
\eeq
It is interesting to underline that this theorem allows to construct 
a general theory of thermodynamics if $\deq$ has a non-trivial symmetry. 
It is easy to see that the constructions of sects. \ref{metrs} and 
\ref{folia} remain unaltered, and also the constructive assumptions 
of sect.\ref{assum} can be trivially generalized [it is sufficient to 
delete ``quasi-homogeneous'' in a1) and a2) and to change a5) into 
``the metrical entropy $S$ satisfies $X\; S = S$].  
The case of quasi-homogeneous symmetry 
is immediately obtained by requiring that each independent variable 
in the thermodynamic space has the same kind of symmetry as the 
potential $S$. More general symmetries can also be introduced. 
Further details will be given elsewhere \cite{belsym}.

%\newpage
\subsection{the quasi-homogeneous case}

The following results are immediate consequences of the 
previous lemmas and theorem in the case of quasi-homogeneous symmetry.\\  
\\
%\newpage
%\noindent
\underline{result 1}:\\ 
\\
{\sl Let 
\beq
\omega=\sum_{i=1}^n\; \omega_i\; dx^i
\eeq
be an exact quasi-homogeneous 
Pfaffian form of degree $r\not = 0$ and weights $(\alpha_1,\ldots,\alpha_n)$;\\
let 
\beq
D\equiv \sum_{i=1}^n\; \alpha_i\; x^i\; \frac{\pa}{\pa x^i} 
\eeq
be the Euler operator;\\
\sl let the symmetry be nontrivial (or transversal), i.e. 
\beq
\omega(D) = i_D\; \omega = \sum_{i=1}^n\; \alpha_i\; \omega_i\; x^i\not 
\equiv 0;
\eeq 
Then, one finds that $\omega=d\phi$ is implemented 
by}  
\beq
\phi \equiv \frac{1}{r}\; \omega(D).
\label{esac}
\eeq
The above result can be used even when some zero weights occur. 
It cannot be used if $r=0$, in which case 
the potential can be found only by quadratures.  
When one considers e.g. $\deq/f$, 
the potential has to be found by integrating the exact form.\\
\\ 
\underline{result 2}:\\ 
\\
{\sl Let $\omega_{(0)}$ be an exact quasi-homogeneous 
Pfaffian form of degree $r=0$;\\ 
let $W$ be the 
associated potential: $d\hat{W}\equiv \omega_{(0)}$.\\
Then 
\beq
d\hat{W}=\frac{dF}{F}
\eeq
where $F$ is a quasi-homogeneous function of degree $q$ with respect 
to the Euler vector field $D$. Moreover, 
\beq
i_D\; \omega_{(0)}=q.
\eeq
As a consequence, 
$q=0$ is allowed if and only if the symmetry is tangent.}\\
\\
Notice that $\hat{W}$ has to be found by quadratures, the 
contraction of the Pfaffian form with the Euler vector field is 
not useful in order to find a potential without explicit integration.  
\\
\\
\underline{result 3}:\\ 
\\
{\sl If $\omega_{(r)}$ is a quasi-homogeneous 
integrable Pfaffian form of degree $r$ defined in a connected, 
simply connected domain, if the symmetry is 
non-trivial and if 
$\mu$ is the corresponding integrating factor $\omega(D)$, 
then 
\beq
d\hat{W}\equiv \frac{\omega_{(r)}}{\mu}=\frac{dF}{F}
\label{dimo}
\eeq
where $F$ is a quasi-homogeneous function of degree one with respect 
to the Euler vector field $D$.}\\
\\
The degree $r$ of $\omega_{(r)}$ can also be zero in the case 
where at least one weight, say $\alpha_1$, is different from zero 
(the would-be intensive variable $\omega_1$ has then weight $-\alpha_1$). 
Herein examples are given where $0<r<\infty$. It would be interesting to 
see if $r\leq 0$ is physically meaningful.\\
Notice that the fact that the case $q=0$ is 
not allowed by the requirement of non-triviality for the symmetry 
can be verified by direct inspection, 
in fact from eqn. (\ref{fome}) it follows
\beq
\pa_i F = \frac{F}{\mu}\; \omega^i \quad \forall i=1,\ldots,n,  
\eeq
thus $D F =0$ becomes
\beq
\sum_{i=1}^{n}\; \alpha_i\; x^i\; \frac{F}{\mu}\; \omega^i = 0,
\eeq
which is not allowed, in fact, because of 
$\mu = \sum_{i=1}^{n}\; \alpha_i\; x^i\; \omega^i$,  
it would imply $F\equiv 0$. It is interesting to note that,  
from a mathematical point of view, 
$\hat{W}$ is an almost quasi-homogeneous function of degree zero 
with first order deficiency index identically equal to one (cf. 
\cite{grudzi}),  
i.e. its behavior under quasi-homogeneous scaling is 
\beq
\hat{W}(\lambda^{\alpha_1} x^1,\ldots,\lambda^{\alpha_n} x^n)= 
\hat{W}(x^1,\ldots,x^n)+\log(\lambda).
\eeq
This property of $\hat{W}$ is to be 
related with the non-triviality of the symmetry, which allows to find 
an integrating factor $\mu\not \equiv 0$.

\subsection{some more results on quasi-homogeneous integrable 
Pfaffian forms}

We have found that, for non-trivial symmetry,  
\beq
\omega_{(r)}=g_{(r-1)}\; dF, 
\label{omef}
\eeq
where $g_{(r-1)}$ is quasi-homogeneous of degree $r-1$. 
If $G$ is a quasi-homogeneous function of degree $q$ such that 
\beq
\omega_{(r)}=g_{(r-q)}\; dG
\label{omeg}
\eeq
then necessarily  
\beq
G =\zeta\; F^{q},
\label{homsup}
\eeq
where $\zeta=$ const. In fact, one has
\beq
\omega_{(r)}=g_{(r-q)}\; dG = g_{(r-1)}\; dF
\eeq
and 
\beq
g_{(r-q)}=\frac{g_{(r-1)}}{dG/dF}.
\eeq
Moreover, the quasi-homogeneity of $G(F)$ implies
\beq
D\; G = q\; G = \frac{dG}{dF}\; F,
\eeq
that is,
\beq
\frac{dG}{dF}=q\; \frac{G}{F},
\eeq
whose solution is (\ref{homsup}). For $q=1$, one obtains that 
the quasi-homogeneous function $F$ of degree one implementing 
(\ref{omef}) is unique apart from a multiplicative constant. 
[Notice that this result can be generalized easily to the 
case of a generic symmetry $X$. By referring to theorem 1, if $G$ satisfies 
$X\; G = q\; G$ and $\omega = b\; dG$, then 
$G=G(F)$, where $F$ is the potential (\ref{exf}), 
and $G=\kappa\; F^q$, where $\kappa$ is a constant].\\
It is evident that, if $D$ is a symmetry generator,  
$D_q = q\; D$ is another symmetry generator for $q\in \RR-\{0\}$.  
It is also interesting that, by changing $q$, the integrating 
factor $f_q$ changes in such a way that $\omega/f_q=d\log(F_q)$, 
where $F_q=F^{1/q}$ is quasi-homogeneous of degree one with 
respect to $D_q$ (i.e., $D_q\; F_q=F_q$). In fact, one has $f_q= q\; f$ 
and 
\beq
\frac{\omega}{f_q}=\frac{1}{q}\; \frac{\omega}{f}= \frac{1}{q}\; d\log(F) = 
d\log(F^{1/q}); 
\eeq
then, $D_q\; F^{1/q}=q\; D\; F^{1/q}=q\; 1/q\; F^{1/q-1}\; D\; F=F^{1/q}$.

\section{$S$ as a function of $T$ in quasi-homogeneous thermodynamics 
and the third law in entropic form}  
\label{thirdap}

We assume that the metrical entropy $S^{\ast}$ is 
quasi-homogeneous of degree one, and that 
$\alpha,\alpha_1,\ldots,\alpha_n$ are the weights of the 
independent variables 
$U^{\ast},X^{1\; \ast},\ldots,X^{n\; \ast}$. 
The temperature 
is then a quasi-homogeneous function of degree $\alpha-1$. Let 
us consider 
$S^{\ast}=S^{\ast}(T^{\ast},X^{1\; \ast},\ldots,X^{n\; \ast})$. 
The entropic form of the third law, for standard thermodynamics, 
states that $S\to S_0$ 
as $T\to 0$, where $S_0$ is a constant which has to be 
zero \cite{belhom}. Planck's restatement of the 
third law $S\to 0$ as $T\to 0$ is mandatory, if the third law holds, 
also in the case of quasi-homogeneous thermodynamics. In fact, if, 
for any fixed value of $X^{1\; \ast},\ldots,X^{n\; \ast})$ 
\beq
\lim_{T^{\ast}\to 0}\; S^{\ast}(T^{\ast},X^{1\; \ast},\ldots,X^{n\; \ast})=
S^{\ast}(0,X^{1\; \ast},\ldots,X^{n\; \ast})\equiv S^{\ast}_0
\eeq
holds, then, from 
\beq
\lim_{T^{\ast}\to 0}\; S^{\ast}(\lambda^{\alpha-1} T^{\ast},
\lambda^{\alpha_1} X^{1\; \ast},\ldots,\lambda^{\alpha_n} X^{n\; \ast})=
\lambda\; \lim_{T^{\ast}\to 0}\; S^{\ast}
(T^{\ast},X^{1\; \ast},\ldots,X^{n\; \ast})
=\lambda\; S^{\ast}_0
\eeq 
and 
\beq
\lim_{T^{\ast}\to 0}\; S^{\ast}(\lambda^{\alpha-1} T^{\ast},
\lambda^{\alpha_1} X^{1\; \ast},\ldots,\lambda^{\alpha_n} X^{n\; \ast})
=S^{\ast}(0,
\lambda^{\alpha_1} X^{1\; \ast},\ldots,\lambda^{\alpha_n} X^{n\; \ast})
=S^{\ast}_0
\eeq 
one finds $S^{\ast}_0=0$. In the above formulas, continuity 
of $S$ at $T=0$ is assumed (otherwise the third law is  
violated \cite{belg30}).\\
We know that black hole thermodynamics violates the 
entropic form of the third law. This behavior is not a general feature 
of quasi-homogeneous thermodynamics. 
We construct a toy-entropy which is both 
quasi-homogeneous and superadditive and implements 
the third law. 
Herein, as in sect. \ref{nonh}, with $x$ we 
indicate collectively the $n$ variables appearing in $S$. 
Our starting point consists in realizing that, given 
two non-negative superadditive functions $g(x),h(x)$, the 
function 
\beq
F(x)\equiv g(x)\; h(x)
\eeq
is superadditive too. In fact, 
\beqnl
F(x_1+x_2)&=& g(x_1+x_2)h(x_1+x_2)
\geq (g(x_1)+g(x_2))(h(x_1)+h(x_2))\\
&=& g(x_1) h(x_1)+g(x_2) h(x_2)+
g(x_1) h(x_2)+ g(x_2) h(x_1)\\
&\geq& g(x_1) h(x_1)+g(x_2) h(x_2)=
F(x_1)+F(x_2).
\eeqnl
Then, let us consider $g=2 \sqrt{U^{\ast} X^{\ast}}$ and $h=X^{\ast\; 2}$. 
Both are 
superadditive functions ($g$ is superadditive because of its homogeneity 
and concavity). Then 
$S^{\ast}\equiv 2 \sqrt{U^{\ast} X^{\ast}}\; X^{\ast\; 2}$ is 
superadditive and quasi-homogeneous. 
For $S^{\ast}$ as a function of $T^{\ast}$, one finds
\beq
S^{\ast}=2\; T^{\ast} X^{\ast\; 5}
\eeq
which satisfies $S^{\ast}\to 0$ as $T^{\ast}\to 0$.\\
It is also interesting to notice that, if $S$ is a strictly 
superadditive function everywhere (i.e., also on the boundary $T=0$) then 
the entropic version of the third law in Planck's form has to be violated; 
in fact, 
if $(U_0,X_0^1,\ldots,X_0^n)$ is such that\\ 
$T(U_0,X_0^1,\ldots,X_0^n)=0$, then 
\beq
S(U+U_0,X^1+X_0^1,\ldots,X^n+X_0^n)>S(U_0,X_0^1,\ldots,X_0^n)+
S(U,X_1,\ldots,X^n)
\eeq
is possible with the strict inequality only if $S(U_0,X_0^1,\ldots,X_0^n)>0$. 
In the case of a self-gravitating system belonging to quasi-homogeneous 
thermodynamics framework, if the energy is 
strictly subadditive everywhere, then the third law cannot hold.\\
As far as mathematical properties of the Pfaffian form $\deq$ 
ensuring the validity of the entropic form of the third law are 
concerned, it can be shown that a quasi-homogeneous $\deq$ which 
is $C^1$ everywhere is a sufficient condition, in analogy 
with standard thermodynamics \cite{belhom,belg30}.  
In fact, a superadditive non-negative 
function $S:{\cal D}\rightarrow \RR_+\cup \{\infty\}$ 
cannot be divergent at a limit point $x_0$ of its domain  
[we mean that all the points $x$ in the domain satisfy $||x||<\infty$, 
where $|| \cdot ||$ stays for the Euclidean norm and that $x_0$ can be 
included in the domain by putting $S(x_0)=\infty$].   
This is evident if one translates rigidly the domain in such a way 
that $x_0\equiv 0$. Then, if $z_0$ is another point in the 
domain of $S$ one has $S(z_0)=S(z_0+0)\geq S(0)+S(z_0)=\infty$. 
Thus, if $S\not \equiv \infty$, then $S$ has to be finite. 
Then $S$ has to be 
finite in the limit as $T\to 0$. As a consequence, if $f$ is the 
integrating factor, one has 
\beq
\frac{\pa f}{\pa U^{\ast}}= 1+S^{\ast}\; \frac{\pa T^{\ast}}{\pa U^{\ast}} 
\eeq
and $\pa T^{\ast}/\pa U^{\ast}=1/C$, where $C$ is the heat capacity at 
fixed parameters $X^{i\; \ast}$, which has to tend to zero if $S^{\ast}$ 
is finite and non-negative as $T^{\ast}\to 0$. Then, if $f$ is everywhere 
$C^1$ (and it is such if $\deq$ is $C^1$ everywhere), 
it holds $S^{\ast}\to 0$ as $T^{\ast}\to 0$. Moreover, it is evident 
from (\ref{qos}) that $\lim_{T^{\ast}\to 0}\; S^{\ast}=0$ holds if and only 
if $\int_{\gamma}\; \delta Q_{rev}/f^{\ast}\to -\infty$ as $T^{\ast}\to 0$ 
whichever path is chosen in approaching $T^{\ast}=0$. Even the latter is 
an extension to the quasi-homogeneous case of the standard case 
\cite{belhom,belg30}. See also \cite{belg31} for further conditions.
%\newpage

\section{Gibbs-Duhem equation revisited}
\label{gdap}

Let us consider a quasi-homogeneous function $g(x^1,\ldots,x^n)$ 
of degree $q\not=0$ 
and of weights $(\alpha_1,\ldots,\alpha_p,0,\ldots,0)$ 
where $1\leq p\leq n$ is an integer and 
the last $n-p$ weights are all zero. 
It satisfies 
\beq
D\; g = q\; g,
\eeq
where
\beq
D=\sum_{i\leq p}\; \alpha_i\; x^i\; \frac{\pa}{\pa x^i}.
\eeq
If one defines
\beq
g_i\equiv \frac{\pa g}{\pa x^i}\quad \forall\; i=1,\ldots,n,
\eeq
and 
\beq
\omega_g\equiv \sum_i\; g_i\; dx^i=dg,
\eeq
one finds 
\beq
g=\frac{1}{q}\; i_D\; \omega_g.
\eeq
This formal manipulation is useful in order to derive the 
Gibbs-Duhem equation. In fact, one has 
\beq
dg=d\; (\frac{1}{q}\; i_D\; \omega_g)=\frac{1}{q}\; \left(
-i_D\; d\omega_g+L_D\; \omega_g \right),
\eeq
where $L_D\; \omega_g=q\; \omega_g$ holds. Moreover, $d\omega_g=0$ 
because $dg=\omega_g$ by construction. The Gibbs-Duhem equation 
is shown to be equivalent to 
\beq
-\frac{1}{q}\; i_D\; d\omega_g =0.
\label{gidu}
\eeq
One can easily see that 
\beq
d\omega_g=\sum_i\; dg_i \wedge dx^i
\eeq
and that 
\beqnl
i_D\; d\omega_g &=& \sum_i\; (i_D\; dg_i)\; dx^i - \sum_i\;  dg_i\; 
(i_D\; dx^i)\cr
&=& \sum_i\; (D\; g_i)\; dx^i - \sum_{i\leq p}\;  dg_i\; (\alpha_i\; x^i)\cr
&=& \sum_{i\leq p}\; (q-\alpha_i)\; g_i\; dx^i
 - \sum_{i\leq p}\;  dg_i\; (\alpha_i\; x^i)+\sum_{p<i\leq n}\; q\; g_i\; dx^i.
\eeqnl
As a consequence, (\ref{gidu}) becomes
\beq
\sum_{i\leq p}\; \frac{\alpha_i}{q}\; (x^i)^{q/\alpha_i}\; 
d\left((x^i)^{1-q/\alpha_i}\; g_i \right)-\sum_{p<i\leq n}\; g_i\; dx^i=0.
\eeq
If none independent variable of weight zero appears, then one finds 
\beq
\sum_i\; \frac{\alpha_i}{q}\; (x^i)^{q/\alpha_i}\; 
d\left((x^i)^{1-q/\alpha_i}\; g_i \right)=0.
\eeq
Let us consider the inverse problem, where one assigns $n$ would-be 
intensive variables $g_i$ and the quasi-homogeneous 1-form of degree 
$\omega_g=\sum_i\; g_i\; dx^i$ such that $L_D\; \omega_g=q\; \omega_g$ 
(one has to require that deg($g_i)=q-\alpha_i$, where the weights 
of the independent variables $x^i$ are the same as above). 
Then one can also define a function 
$g\equiv i_D \omega_g/q$. From $dg=(-i_D\; d\omega_g)/q+\omega_g$ follows 
that $\omega_g$ is closed if $i_D\; d\omega_g=0$, i.e., if the 
Gibbs-Duhem equation is satisfied.

We limit ourselves to point out that 
more general Gibbs-Duhem equations are obtained in the generic 
symmetric case \cite{belsym}. The generalization of (\ref{giduw}) is
\beq
(i_X\; \deq)\; d \log \left( \frac{1}{T} \right)-i_X\; d\; \deq+ 
(L_X-1)\; \deq=0.
\eeq

%\newpageq

\section{quasi-homogeneity (scaling) cannot be generalized}
\label{qoap}

We show that, if $f(x^1,\ldots,x^n)$ is a $C^1$ function on a open 
connected set which satisfies the following identity:
\beq
f(g_1 (\lambda)\; x^1,\ldots,g_n (\lambda)\; x^n)= 
g (\lambda)\; f(x^1,\ldots,x^n)
\eeq
and the functions $g (\lambda), g_i (\lambda)$ for $i=1,\ldots,n$ are 
positive definite and invertible, then $f(x^1,\ldots,x^n)$ is necessarily 
quasi-homogeneous. The above transformation is meant to be obtained 
by a ``generalized similarity transformation'' which carries 
$x^1,\ldots,x^n$ into 
$g_1 (\lambda)\; x^1,\ldots,g_n (\lambda)\; x^n$.  
Let us consider $u \equiv g(\lambda)$ and the 
inverse $\lambda = g^{-1} (u)$. Define
\beq
h_i (u) \equiv g_i (g^{-1}(u))\quad \forall  i=1,\ldots,n.
\eeq
It is also useful to define 
\beq
u\equiv e^s.
\eeq
Then we obtain 
\beq
f(h_1 (e^s)\; x^1,\ldots,h_n (e^s)\; x^n)= 
e^s\; f(x^1,\ldots,x^n).
\eeq
Note that, for $s=0$, one finds
\beq
f(h_1 (1)\; x^1,\ldots,h_n (1)\; x^n)= f(x^1,\ldots,x^n).
\eeq
This a priori does not imply that $h_i (1)=1\quad \forall i=1,\ldots,n$, 
but the one-parameter $s\in \RR_+$ ``generalized similarity transformation'' 
${\cal T}_s:\; (x^1,\ldots,x^n) \to  
(h_1 (e^s)\; x^1,\ldots,h_n (e^s)\; x^n)$ is consistently defined if 
${\cal T}_{s=0}$ is the identity, i.e. if 
$h_i (1)=1\quad \forall i=1,\ldots,n$.\\
We introduce the auxiliary function 
\beq
F(s;x^1,\ldots,x^n) \equiv e^{-s}\; f(h_1 (e^s)\; x^1,\ldots,h_n (e^s)\; x^n).
\eeq
It is such that $\pa_s F = 0$, i.e.
\beq
F=\sum_{i=1}^{n}\; e^s\; h^{\prime}_i (e^s)\; x^i\; 
\left(\frac{\pa F}{\pa x^i}\right) (h_1 (e^s)\; x^1,\ldots,h_n (e^s)\; x^n),
\eeq
where $h^{\prime}_i (u)\equiv dh_i/du$. 
For $s=0$ one obtains
\beq
f(x^1,\ldots,x^n)=
\sum_{i=1}^{n}\; h^{\prime}_i (1)\; x^i\; 
\left(\frac{\pa f}{\pa x^i}\right)(x^1,\ldots,x^n).
\eeq
Thus, $f(x^1,\ldots,x^n)$ is a quasi-homogeneous 
function of degree one and weights 
$(h^{\prime}_1 (1),\ldots,h^{\prime}_n (1))\equiv 
(\alpha_1,\ldots,\alpha_n)$. 
As a consequence, one has 
\beq
f(e^{\alpha_1\; s} x^1,\ldots,e^{\alpha_n\; s} x^n)
=e^s\;  f(x^1,\ldots,x^n), 
\eeq
i.e.
\beq
h_i (e^s)=e^{\alpha_i\; s}\quad \forall i=1,\ldots,n.
\eeq
By resorting the original dependence on $\lambda$, one 
finds
\beq
f((g(\lambda))^{\alpha_1} x^1,\ldots,(g(\lambda))^{\alpha_n} x^n)
=g(\lambda)\;  f(x^1,\ldots,x^n), 
\eeq
i.e.
\beq
g_i(\lambda)=(g(\lambda))^{\alpha_i}\quad \forall i=1,\ldots,n.
\eeq
If some $g_i=1$ above, then one simply finds that the variable 
$x^i$ is quasi-homogeneous of degree zero, i.e., it has weight 
zero. If instead $g(\lambda)=1$, then $f$ is quasi-homogeneous 
of degree zero. The proof is straightforward. Let $\lambda_0$ be 
such that $g_i(\lambda_0)=1\quad \forall i=1,\ldots,n$. 
We have that 
\beq
\frac{\pa f}{\pa \lambda}=0=
\sum_{i=1}^{n}\; g^{\prime}_i (\lambda)\; x^i\; 
\left(\frac{\pa f}{\pa x^i}\right) (g_1 (\lambda) x^1,\ldots,
g_n (\lambda) x^n).
\eeq
By setting $\lambda=\lambda_0$, one finds 
\beq
\sum_{i=1}^{n}\; g^{\prime}_i (\lambda_0)\; x^i\; 
\left(\frac{\pa f}{\pa x^i}\right) (x^1,\ldots,x^n),
\eeq
i.e., $f$ is quasi-homogeneous of degree zero and weights 
$(g^{\prime}_1 (\lambda_0),\ldots,g^{\prime}_n (\lambda_0))$.

\section{scaling and asymptotics}
\label{asyap}

Let us assume that $f(x)$ is a  function of 
$n$ variables collectively indicated with $x$. Let the domain 
of $f$ be  invariant under 
quasi-homogeneous transformations. 
Let $\{ {\cal T}_{\lambda} \}$ be a one-parameter quasi-homogeneous 
transformation. If we require that there exist a positive continuous 
function $\rho (\lambda)$ and a continuous function $g(x)\not \equiv 0$ 
such that $g$ is an asymptotic of $f$ in the following sense: 
\beq
\lim_{\lambda \to \infty}\; \frac{1}{\rho (\lambda)}\; 
f ({\cal T}_{\lambda}\; x)=g(x),
\eeq
where the limit is assumed to exist. 
Then:\\
a) the function $\rho (\lambda)$ is a regularly varying function, i.e. 
it satisfies for all $a>0$ 
\beq
\lim_{\lambda \to \infty}\; \frac{\rho (\lambda\; a)}{\rho (\lambda)}\equiv 
C(a),
\eeq
where $C(a) C(b)=C(a\; b)$, i.e. $C(a)=a^{\gamma}$ 
for some $\gamma\in \RR$;  $\gamma$ is also defined the order of the 
regularly varying function $\rho (\lambda)$;\\
b) the function satisfies the ``homogeneity'' relation 
\beq
g ({\cal T}_{\mu}\; x)=\mu^{\gamma}\; g(x),
\eeq
i.e. $g(x)$ is quasi-homogeneous of degree $\gamma$.\\
\\
\underline{proof}: for each real $s>0$ it holds
\beq
\lim_{\lambda \to \infty}\; \frac{1}{\rho (\lambda\; s)}\; 
f ({\cal T}_{\lambda\; s}\; x)=g(x),
\eeq
and, because of ${\cal T}_{\lambda\; s}={\cal T}_{\lambda}\; {\cal T}_{s}$, 
\beq
\lim_{\lambda \to \infty}\; \frac{1}{\rho (\lambda)}\; 
f ({\cal T}_{\lambda}\; ({\cal T}_{s}\; x))=g({\cal T}_{s}\; x).
\eeq
As a consequence, 
\beq
\lim_{\lambda \to \infty}\; 
\frac{\rho (\lambda)}{\rho (\lambda\; s)}\; 
\frac{f ({\cal T}_{\lambda\; s}\; x)}
{f ({\cal T}_{\lambda}\; ({\cal T}_{s}\; x))} =   
\frac{g(x)}{g({\cal T}_{s}\; x)}=
\lim_{\lambda \to \infty}\; \frac{\rho (\lambda)}{\rho (\lambda\; s)}.
\eeq
This is possible only if the following conditions 
\beq
\lim_{\lambda \to \infty}\; \frac{\rho (\lambda\; s)}{\rho (\lambda)}=C(s)
\eeq
and 
\beq
g({\cal T}_{s}\; x)=C(s)\; g(x)
\eeq
are both satisfied. Moreover, it is easy to see that it holds 
\beq
C(s\; t)=C(s)\; C(t),
\eeq
which implies that there exists a real number $\gamma$ such that 
$C(s)=s^{\gamma}$. Thus, one obtains 
\beq
g({\cal T}_{s}\; x)=s^{\gamma}\; g(x),
\eeq
which concludes the proof. \quad \quad  $\square$\\
Cf. also Ref. \cite{vladimirov}.

%\end{multicols}
\end{document}